\documentclass[a4paper,fleqn,usenatbib,twocolumn]{mnras}

\usepackage{newtxtext,newtxmath}

\usepackage[T1]{fontenc}
\usepackage{ae,aecompl}

\usepackage{graphicx}	
\usepackage{amsmath}	
\usepackage{amssymb}	
\usepackage[usenames,dvipsnames]{color} 
\usepackage{pdflscape}

\definecolor{purple}{rgb}{0.5,0,0.5}

\def \figwidth {0.47\textwidth}

\newcommand{\brvs}{Br\"unt-V\"ais\"al\"a}

\title[Turbulence Closure for Mixing Length Theories]{Turbulence Closure for Mixing Length Theories}

\author[Adam S. Jermyn et al.]{
Adam S. Jermyn,$^{1}$\thanks{E-mail: adamjermyn@gmail.com}
Pierre Lesaffre,$^{2}$
Christopher A. Tout$^{1}$
and Shashikumar M. Chitre$^{1,3}$
\\
$^{1}$Institute of Astronomy, University of Cambridge, Madingley Rd, Cambridge CB3 0HA, UK\\
$^{2}$\'{E}cole Normale Sup\'{e}rieure 24 rue Lhomond, 75231 Paris, France\\
$^{3}$Centre for Basic Sciences, University of Mumbai, India
}

\date{Accepted XXX. Received YYY; in original form ZZZ}

\pubyear{2017}

\begin{document}
\label{firstpage}
\pagerange{\pageref{firstpage}--\pageref{lastpage}}
\maketitle

\begin{abstract}

We present an approach to turbulence closure based on mixing length theory with three-dimensional fluctuations against a two-dimensional background.
This model is intended to be rapidly computable for implementation in stellar evolution software and to capture a wide range of relevant phenomena with just a single free parameter, namely the mixing length.
We incorporate magnetic, rotational, baroclinic and buoyancy effects exactly within the formalism of linear growth theories with nonlinear decay.
We treat differential rotation effects perturbatively in the corotating frame using a novel controlled approximation which matches the time evolution of the reference frame to arbitrary order.
We then implement this model in an efficient open source code and discuss the resulting turbulent stresses and transport coefficients.
We demonstrate that this model exhibits convective, baroclinic and shear instabilities as well as the magnetorotational instability (MRI).
It also exhibits non-linear saturation behaviour, and we use this to extract the asymptotic scaling of various transport coefficients in physically interesting limits.

\end{abstract}

\begin{keywords}
convection - Stars; rotation - Stars; interiors - Stars: evolution
\end{keywords}



\section{Introduction}

An understanding of turbulent transport and stresses remains one of the major outstanding problems in the astrophysics of fluids.
While many pieces of this puzzle are understood in broad strokes, the nature of this problem is such that the details are almost as important as the big picture.
The magnetorotational instability (MRI), for instance, is understood conceptually but making predictions which match observed accretion discs is a persistent problem\ \citep{2015ApJ...802..139M}.
Similarly the solar differential rotation is understood to arise from turbulent stresses but precisely how this works and in balance with what other forces remains uncertain\ \citep{1998ApJ...505..390S}.

Significant progress has indeed been made with three-dimensional turbulence simulations~\citep[for examples see][]{2013JCoPh.243..269L, 2014MNRAS.441.3177M, 2016MNRAS.457..857S} but these are generally relevant only on short timescales and in small volumes.
Performing so-called global simulations over large times and distances requires a turbulence closure model to substitute for resolution at small scales\ \citep{LAUNDER1974269,1994ApJ...428..729C}.

At the other extreme models of stellar evolution generally assume extremely simple analytical transport coefficients to overcome the tremendous gap between turbulent timescales of minutes and nuclear timescales of millions of years\ \citep{1995A&A...299...84M}.
A variety of such approaches have been developed.
For instance the mixing length theory of~\citet{1958ZA.....46..108B} provided a closure of convection.
This was then put on firmer theoretical ground by~\citet{1977LNP....71...15G,2012ISRAA2012E...2G} and extended to include additional phenomena~\citep{2011IAUS..271..397S,2013MNRAS.431.2200L}.
\citet{1986GApFD..35...93K} introduced an entirely different closure formalism, arriving at an expression for the so-called $\Lambda$-effect~\citep{1987GApFD..38..273K}, and later incorporating it under the $\alpha-\Lambda$ formalism with Rudiger~\citep{1993A&A...276...96K}.
What these formalisms have in common is a minimal set of free parameters: the mixing length formalism has just the mixing length, and the formalism of~\citet{1993A&A...276...96K} has just the anisotropy parameter.

Another set of models has arisen which aims to reproduce higher-order moments of the turbulent fields.
This increases the number of free parameters and a number of approaches have been developed to deal with this.
For instance~\citet{0004-637X-837-2-133} and~\citet{2010MNRAS.407.2451G} fit their free parameters against small-scale simulations while \citet{1997ApJ...482..827C} fits his against experimental results.
In addition there are models, such as that of~\citet{1994ApJ...428..729C}, which fix at least some free parameters by introducing new assumptions, in this case regarding the various relevant time-scales.
Regardless of the details of how they close the equations of turbulent moments, models of this sort generally take the form of physically motivated analytic expressions which provide ready access to scaling laws.
Their free parameters then serve to better their agreement with data, at the cost of being less straightforwardly interpreted and extended.

The availability of growing computational resources in recent years has provided a new niche in this landscape in the form of computational closure models.
These are models which do not seek analytic solutions but which are nonetheless distinct from attempts to simulate turbulence in all its detail.
Some may introduce new dynamical fields, as in the $k-\epsilon$ model\ \citep{LAUNDER1974269}, while others invoke effective theories of small-scale motion\ \citep{1986A&A...168...89C}.
The latter kind are essentially renormalized theories which accept the cost of having to numerically accommodate complex behaviour in exchange for more precision over a wider variety of phenomena.
Combined with perturbation theory this approach represents a tunable middle-ground between expensive simulations and simple analytic models, allowing the computational cost to be
traded off against fidelity to suit the problem at hand.
The model we present here is in this spirit.

We construct a mixing-length theory which incorporates three-dimensional fluctuations against a two-dimensional axisymmetric background.
This is done by treating each mode as growing with its linear growth rate before saturating at an amplitude set by the turbulent cascade\ \citep{2013MNRAS.431.2200L}.
Beyond this the motion in each mode is taken to be uncorrelated.
We treat the geometry of the flow in full generality, allowing for baroclinic effects as well as magnetism and rotational shears.
To incorporate differential rotation we use a time-dependent sheared coordinate system\ \citep{2012MNRAS.426.1546B}.
In this frame there is a continual flow of modes across Fourier space, lending a time dependence to growth rates.
Corrections to saturation amplitudes owing to this flow are incorporated perturbatively with the time derivatives of the growth rate.

In Section\ \ref{sec:closure} we describe our closure framework in more detail, paying particular attention to the choice of mixing length.
We then develop a perturbative approach for correcting the saturation amplitude in Section\ \ref{sec:perturb}.
In section\ \ref{sec:motion} we introduce the sheared coordinate system and the linearised equations of motion.
Finally in Section\ \ref{sec:results} we show results from our theory, including calculations for the solar convection zone and accretion discs.

The software implementing our model is open source and available under a GPLv3 license.
Details of the implementation are given in Appendix\ \ref{appen:software}.
Tabulated transport coefficients produced by the code are also available under the same license and both may be found at \url{github.com/adamjermyn/Mixer}.

\section{Closure Formalism}
\label{sec:closure}

Turbulent phenomena generically exhibit a cascade of energy between large and small scales\ \citep{rgfluids2, ISI:000274107900015}.
With some notable exceptions\ \citep{2007JAtS...64.3312S} this cascade begins at a large scale $L_0$ set by the overall structure of the fluid flow and ends at an extremely small scale $L_\nu$ related to the microscopic viscosity.
Between these scales, yet far from each of them, lies the so-called inertial range where the fluid flow is scale-free\ \citep{1941DoSSR..32...16K}.
In this range all correlations of the turbulent motion obey simple power laws.

This statement was originally proved by\ \citet{1941DoSSR..32...16K} for isotropic turbulence.
It was later found to be a broader consequence of the renormalizability of the Navier-Stokes equation\ \citep{1986JSCom...1....3Y, PhysRevA.41.3129} and consequently holds quite generally.
This means that there is a single relevant scale $L_0$ for a given turbulent flow which fully characterises the turbulence as seen by measurements performed over length scales $L \gg L_0$.
This is the modern interpretation and justification of the original mixing length hypothesis, which asserts that turbulent fluctuations on scales $L \ll L_0$ are not dynamically coupled to the large-scale ($L \gg L_0$) flow properties\ \citep{1958ZA.....46..108B}.

The scale-free nature of turbulence in the inertial range means that modes of significantly different wavevectors are uncorrelated.
A natural extension of this is to assume that all modes of distinct wavevectors are at least approximately uncorrelated.
That is, we assume that
\begin{align}
	\langle \tilde{\boldsymbol{v}}_{\boldsymbol{k}} \otimes \tilde{\boldsymbol{v}}^*_{\boldsymbol{k}'} \rangle &= (2\pi)^3 \delta^3 (\boldsymbol{k}-\boldsymbol{k}') \mathsf{V}_{\boldsymbol{k}},
	\label{eq:correl1}
\end{align}
where $v$ is the velocity, $\otimes$ denotes the outer product, $\langle ... \rangle$ denotes the time-averaged expectation, $\tilde{\boldsymbol{v}}_{\boldsymbol{k}}$ is the amplitude of the Fourier mode with wavevector $\boldsymbol{k}$ and $\mathsf{V}_{\boldsymbol{k}}$ is the tensor specifying how different components of the same mode are correlated with one another.
It is crucial to notice that the quantity $\mathsf{V}_{\boldsymbol{k}}$ is also the Reynolds stress of mode $\boldsymbol{k}$.
This, and several other closely related quantities, are ultimately what we seek.
These two-point correlation functions suffice to characterise not only the stresses but also all higher-order correlations through Wick's theorem and perturbation theory\ \citep{PhysRev.80.268, doi:10.109}.

To determine $\mathsf{V}_{\boldsymbol{k}}$ we begin by writing the linearised equations of motion as
\begin{align}
	\partial_t \boldsymbol{v}(\boldsymbol{r}) = \mathcal{L}\left[\boldsymbol{v}(\boldsymbol{r}), \partial_i \boldsymbol{v}, \partial_i \partial_j \boldsymbol{v}, ..., \boldsymbol{r}, t\right],
\end{align}
where $\mathcal{L}$ is a linear operator of its first argument and $\boldsymbol{v}$ is the fluctuating part of the velocity field.
In principle we can work with this operator, though the derivatives of the velocity field make it highly inconvenient.
Fortunately at short length scales the operator $\mathcal{L}$ may be treated as translation-invariant and so we may compute a Fourier transform in $\boldsymbol{r}$ without coupling different modes.
This gives
\begin{align}
	\frac{d\tilde{\boldsymbol{v}}_{\boldsymbol{k}}}{dt} = \tilde{\mathcal{L}}\left[\tilde{\boldsymbol{v}}_{\boldsymbol{k}}, \boldsymbol{k}, t\right].
\end{align}
The modes are decoupled in this regime so $\tilde{\mathcal{L}}$ can be represented by a matrix $\mathsf{L}$, and we write
\begin{align}
	\frac{d\tilde{\boldsymbol{v}}_{\boldsymbol{k}}}{dt} = \mathsf{L}(\boldsymbol{k},t) \boldsymbol{v}_{\boldsymbol{k}}.
	\label{eq:linear}
\end{align}

When $\mathsf{L}$ is independent of $t$ equation\ \eqref{eq:linear} is straightforward to solve and gives us
\begin{align}
	\frac{d\tilde{\boldsymbol{v}}_{\boldsymbol{k}}}{dt} = \sum_i v_{0,i} \hat{v}_{\boldsymbol{k},i} e^{\lambda_i t},
	\label{eq:sol}
\end{align}
where $v_{0,i}$ are the initial mode amplitudes and $\hat{v}_{\boldsymbol{k},i}$ and $\lambda_i$ are respectively the normalised right eigenvectors and eigenvalues of $\mathsf{L}$.
The vectors $\hat{v}_{\boldsymbol{k},i}$ then specify the modes of the system at a given wavevector.

If the eigenvalues are not precisely degenerate then modes which begin in phase rapidly become uncorrelated and we may extend equation\ \eqref{eq:correl1} to the modes at each wavevector and write
\begin{align}
	\langle \tilde{\boldsymbol{v}}_{\boldsymbol{k},i} \otimes \tilde{\boldsymbol{v}}^*_{\boldsymbol{k}',j} \rangle &= (2\pi)^3 \delta^3 (\boldsymbol{k}-\boldsymbol{k}') \delta_{ij} \mathsf{V}_{\boldsymbol{k},i}.
	\label{eq:correl2}
\end{align}
This result holds even when modes are degenerate.
Because the time evolution of the Navier-Stokes equation is deterministic, the expectation $\langle ... \rangle$ represents a sum over initial conditions.
In this sum all relative phases between the modes are explored, so even degenerate modes become uncorrelated.

Inserting equation\ \eqref{eq:sol} into equation\ \eqref{eq:correl2} and summing over $j$ and integrating over $\boldsymbol{k}$ gives us
\begin{align}
	\mathsf{V}_{\boldsymbol{k},i} &= \hat{v}_{\boldsymbol{k},i} \otimes\hat{v}_{\boldsymbol{k},i} \langle |v_{0,i}|^2 \exp\left[2 t \Re\left[\lambda_i\right]\right] \rangle.
	\label{eq:correl3}
\end{align}
Generally some $\lambda_i$ have positive real parts and so in a long-term expectation this exponential diverges.
Indeed it turns out that these growing modes are precisely those which matter!
What happens of course is just that these modes eventually reach amplitudes where the linear approximation fails.
By assumption the system is stable over long times relative to the turbulent scale so this must result in these modes saturating.
This has been variously described as mode crashing or the action of parasitic modes\ \citep{2009ApJ...698L..72P, 2009MNRAS.396..779L} but, regardless of the mechanism, it simply means that these modes exit the linear regime and find their growth impeded.

To complete the closure we must find the saturation amplitude.
Relying again on the scale-free nature of turbulence we note that this must be a power law in $k$.
That is,
\begin{align}
	\langle \tilde{v}_{\boldsymbol{k}, i}^2 \rangle = \mathrm{Tr}\left[\mathsf{V}_{\boldsymbol{k}, i}\right] = \frac{A}{M} \left(\frac{k_0}{k}\right)^{2n},
	\label{eq:amp}
\end{align}
where $A$ depends on the large scale properties of the flow but is independent of $\boldsymbol{k}$, $M$ is the number of modes per wavevector and $n$ is the index of the turbulence.
Following \citet{1941DoSSR..30..301K} we choose $n=11/6$ in our model.
Appendix\ \ref{appen:n} contains a detailed discussion of this choice.

The wavenumber $k_0$ is just that of the characteristic scale, and is given by
\begin{align}
	k_0 = \frac{2\pi}{L_0}.
	\label{eq:k0}
\end{align}
Replacing the divergent expression in equation\ \eqref{eq:correl3} with this amplitude we find
\begin{align}
	\mathsf{V}_{\boldsymbol{k},i} &= \frac{A}{M} \left(\frac{k_0}{k}\right)^{2n} \hat{v}_{\boldsymbol{k},i} \otimes\hat{v}_{\boldsymbol{k},i}.
	\label{eq:correl4}
\end{align}

It only remains to determine $A$.
To do this we note that there is one characteristic length scale $L_0$ and one characteristic timescale, the growth rate $\Re\left[\lambda_i\right]$ of the mode.
Because $A$ has dimensions of velocity squared we find
\begin{align}
	\mathsf{V}_{\boldsymbol{k},i} &= \frac{c}{M} L_0^2 \Re\left[\lambda_i\right]^2 \left(\frac{k_0}{k}\right)^{2n} \hat{v}_{\boldsymbol{k},i} \otimes\hat{v}_{\boldsymbol{k},i},
	\label{eq:correl5}
\end{align}
where $c$ is a dimensionless constant of order unity.
This constant, known as the mixing length parameter, varies from theory to theory, so for clarity we set $c=1$ in this work but this degree of freedom is important to note when comparing between models.
In effect what we have done is incorporate the non-linearity of turbulence by means of the spectrum while using linear growth rates to set the characteristic scale.
In practice the spectrum only acts to provide a convergent measure over modes (see Appendix~\ref{appen:n} for further discussion), and it is the growth rate and the modes themselves that yield the anisotropies and other phenomena of interest.
This is closely related to the approaches of~\citet{2013MNRAS.431.2200L} and~\citet{1986A&A...168...89C}.

This prescription is easily extended in cases where there are additional dynamical fields, such as the turbulent displacement or a fluctuating magnetic field.
The additional fields are simply incorporated into the vector describing the state and $M$ is increased accordingly.
We can continue to use equation\ \eqref{eq:amp} to fix the amplitude of the entire mode against that of the velocity as long as we know the turbulent index $n$.

Up to this point this prescription is mathematically identical to that of\ \citet{2013MNRAS.431.2200L}, with the exception that we define the mixing wave vector as in equation\ \eqref{eq:k0} while they use $\pi/L_0$ instead.
In the next section we introduce perturbative corrections to this model to capture a wider variety of phenomena.

\section{Perturbative Corrections}
\label{sec:perturb}

Now consider the case where the matrix $\mathsf{L}$ is time-dependent.
Most of our reasoning about the behaviour of modes from the previous section still holds but, because the eigenvectors are time-dependent, we no longer have a well-defined notion of a mode as a long-running solution to the equations of motion.
When the time dependence is periodic Floquet theory applies~\citep{Floquet1883}, but in the cases of interest the time dependence is aperiodic.
To recover modes when the time evolution matrix itself evolves and does so aperiodically we begin by expanding as
\begin{equation}
	\mathsf{L}(t) = \mathsf{L}(0) + t \frac{d\mathsf{L}}{dt} + \frac{1}{2}t^2 \frac{d^2\mathsf{L}}{dt^2} + ...\,.
\end{equation}
This series can be truncated to produce an approximation of $\mathsf{L}$ which is accurate in a certain window around $t=0$.

We may likewise write the velocity at a given wavevector as
\begin{equation}
	\tilde{\boldsymbol{v}}_{\boldsymbol{k}}(t) = \tilde{\boldsymbol{v}}_{\boldsymbol{k}}(0) + t\left.\frac{d\tilde{\boldsymbol{v}}_{\boldsymbol{k}}}{dt}\right|_0 + \frac{1}{2}t^2\left.\frac{d^2\tilde{\boldsymbol{v}}_{\boldsymbol{k}}(t)}{dt^2}\right|_0 + ...\,.
\end{equation}
This suggests defining a new vector
\begin{equation}
	\Phi_{\boldsymbol{k}}(t) \equiv \left\{\tilde{\boldsymbol{v}}_{\boldsymbol{k}}, \frac{d\tilde{\boldsymbol{v}}_{\boldsymbol{k}}}{dt}, \frac{d^2\tilde{\boldsymbol{v}}_{\boldsymbol{k}}}{dt^2}, ...\right\},
\end{equation}
which, in principle, encodes the full time evolution of the velocity field.
This vector evolves according to
\begin{equation}
	\frac{d\Phi_{\boldsymbol{k}}}{dt} = \mathsf{A} \Phi_{\boldsymbol{k}}
\label{eq:dt}
\end{equation}
where $\mathsf{A}$ is formed of blocks given by
\begin{align}
	\mathsf{A}_{ij} = {i \choose j}\frac{d^{i-j}}{dt^{i-j}} \mathsf{L}.
\end{align}
By definition though we also have
\begin{align}
		\frac{d\Phi_{\boldsymbol{k},i}}{dt} = \Phi_{\boldsymbol{k},i+1},
		\label{eq:constraint}
\end{align}
where $\Phi_{\boldsymbol{k},0} = \tilde{\boldsymbol{v}}_{\boldsymbol{k}}$, $\Phi_{\boldsymbol{k},1} = d\tilde{\boldsymbol{v}}_{\boldsymbol{k}}/dt$ and so on.
Thus we are searching for a simultaneous solution of equations\ \eqref{eq:dt} and\ \eqref{eq:constraint}.

In order to close the system we must truncate it at some finite order $N$.
Doing so makes the assumption that the behaviour of the system at all greater $N$ is known.
Inspired by the solution for time-independent $\mathsf{L}$, we try an exponential behaviour.
This truncates equation\ \eqref{eq:constraint} such that it applies only to $i < N-1$ and means that we are searching for vectors with
\begin{align}
	\left(A \Phi_{\boldsymbol{k}}\right)_{N-1} = \lambda \Phi_{\boldsymbol{k},N-1}
\end{align}
and
\begin{align}
	\Phi_{\boldsymbol{k},i+1} = (\mathsf{A} \Phi_{\boldsymbol{k}})_i, i < N-1.
	\label{eq:constraint2}
\end{align}
These equations are most straightforwardly written as a general eigensystem and this has the advantage of restricting the dimension of the linear space to just those states obeying the constraint.
This is possible because both $\mathsf{A}$ and the constraint are lower-triangular in the same basis, and so each row may be substituted into the next, leading to an eigenproblem of the form
\begin{align}
	\mathsf{Q} \Phi_{\boldsymbol{k,0}} = \lambda \mathsf{W} \Phi_{\boldsymbol{k,0}},
	\label{eq:generaleigen}
\end{align}
where $\mathsf{Q}$ and $\mathsf{W}$ are matrices acting only on the $0$-block.
For example, in the case where $N=2$, our equations are
\begin{align}
	\Phi_{\boldsymbol{k},1} &= \mathsf{M} \Phi_{\boldsymbol{k},0}\\
\intertext{and}
	\mathsf{M} \Phi_{\boldsymbol{k},1} + \dot{\mathsf{M}} \Phi_{\boldsymbol{k},0} &= \lambda \Phi_{\boldsymbol{k},1},	
\end{align}
which may be put in the form of equation\ \eqref{eq:generaleigen} with
\begin{align}
	\mathsf{Q} &= \mathsf{M}^2 + \dot{\mathsf{M}}\\
\intertext{and}
	\mathsf{W} &= \mathsf{M}.
\end{align}
The eigenvectors of this system are solutions of the original equation\ \eqref{eq:linear} because if $\psi^i_{\boldsymbol{k}}$ is such an eigenvector then
\begin{equation}
	\tilde{\boldsymbol{v}}_{\boldsymbol{k},i}(t) \equiv \sum_{j=0}^{N} \frac{t^j}{j!}\psi^i_{j} 
\end{equation}
solves
\begin{equation}
	\frac{d\tilde{\boldsymbol{v}}_{\boldsymbol{k},i}}{dt} = \mathsf{L}(t) \tilde{\boldsymbol{v}}_{\boldsymbol{k},i}(t)
\end{equation}
over the time window for which $\mathsf{L}$ is well-approximated at $N$-th order.
As a result we say that $\phi_i(t)$ are the instantaneous modes of the system at $N$-th order and use them and in equation\ \eqref{eq:correl5}.
In place of the eigenvalue we use the instantaneous growth rate of the velocity, which is given by
\begin{align}
	g \equiv \frac{1}{2}\frac{d v^2}{dt} = \frac{\Re\left(\Phi_{\boldsymbol{k},0} \cdot \Phi_{\boldsymbol{k},1}\right)}{|\Phi_{\boldsymbol{k},0}|^2}.
\end{align}
This approximation is controlled in the sense that so long as $\mathsf{L}(t)$ converges as $N$ grows, so does the inferred velocity history.
In this work we present results with $N=2$ so that $\mathsf{A}$ involves both $\mathsf{L}$ and $\dot{\mathsf{L}}$.
We leave the exploration of larger $N$ to later work.

\section{Equations of Motion}
\label{sec:motion}

We now specialise to the case of an ideal gas obeying the ideal MHD equations.
This section largely follows the derivation of\ \citet{2012MNRAS.426.1546B} so we present only the pieces necessary to understand later parts of this work as well as the few places where our derivation diverges from theirs.

We take the background to be axisymmetric, the fluctuations to be adiabatic and we work in cylindrical coordinates.
We neglect both the microscopic viscosity and the microscopic thermal diffusivity because these are both negligible in most circumstances in stellar physics\footnote{It would not be difficult, however, to incorporate them into this framework at a later date.}. 
Because our closure model treats turbulent properties as local, we compute all background quantities at a reference point $\boldsymbol{r}_0$.
Relative to this point we define the Lagrangian separation $\boldsymbol{\delta r}$ and velocity $\boldsymbol{\delta v}$ equivalent to $\boldsymbol{\xi}$ and $D\boldsymbol{\xi}/Dt$ of\ \citet{2012MNRAS.426.1546B}.
In addition we take the Boussinesq approximation that density variations are ignored except in terms involving gravitational acceleration.
With the above definitions the continuity equation may be written as
\begin{equation}
	\nabla\cdot\boldsymbol{\delta r} = 0.
	\label{eq:bous}
\end{equation}

In a fixed coordinate system differential rotation is difficult to analyze so we make two reference frame changes.
First we switch from an inertial frame to one rotating at
\begin{equation}
	\Omega_0 \equiv \Omega(\boldsymbol{r}_0).
\end{equation}
Secondly we make a formal change of coordinates
\begin{equation}
\phi \rightarrow \phi - t\delta\boldsymbol{r}\cdot\nabla\Omega
\end{equation}
without altering the corresponding unit vectors.
Under this last change the gradient transforms as
\begin{align}
\nabla &\rightarrow \nabla - t(\nabla\Omega)\partial_{\phi}.
\label{eq:grad}
\end{align}
Because the operator $\mathcal{L}$ is most easily expressed in Fourier space we define the transformed wavevector as
\begin{equation}
	\boldsymbol{q} \equiv \boldsymbol{k} - t k_\phi R \nabla \Omega.
\end{equation}
With this the transformed MHD and Navier-Stokes equations may be written as
\begin{equation}
	\boldsymbol{\delta \tilde{B}} = \boldsymbol{B} \cdot \boldsymbol{q} \boldsymbol{\delta \tilde{r}}
	\label{eq:mhd}
\end{equation}
and
\begin{equation}
\partial_t \delta\tilde{\boldsymbol{v}} + 2\boldsymbol{\Omega}\times\delta\tilde{\boldsymbol{v}}+ \hat{R} R \delta\tilde{\boldsymbol{r}}\cdot\nabla\Omega^2 - \frac{1}{\gamma \rho}\left(\boldsymbol{\delta \tilde{r}}\cdot\nabla\sigma\right)\nabla\cdot\boldsymbol{\Pi} + \frac{i}{\rho}\boldsymbol{q}\cdot\boldsymbol{\delta \tilde{\Pi}}=0,
\label{eq:momentum}
\end{equation}
where $\sigma$ is the specific entropy and
\begin{equation}
	\boldsymbol{\Pi} \equiv p \mathsf{I} - \frac{1}{\mu_0}\left(\boldsymbol{B}\otimes \boldsymbol{B} - \frac{1}{2}B^2 \mathsf{I}\right)
\end{equation}
is the pressure tensor with $\mathsf{I}$ the identity matrix.
All quantities prefixed with $\delta$ are fluctuating, a tilde denotes the Fourier transformed function, and all other quantities are background fields evaluated at $\boldsymbol{r}_0$.
It is straightforward to see that this is the same equation as that derived by \citet{2012MNRAS.426.1546B} once the appropriate relations for the pressure and magnetic force are substituted.

The fluctuation in the pressure tensor may be written as
\begin{equation}
	\boldsymbol{\delta \Pi} = \delta p \mathsf{I} - \frac{1}{\mu_0}\left(\boldsymbol{B}\otimes \boldsymbol{\delta B} + \boldsymbol{\delta B}\otimes \boldsymbol{B} - \mathsf{I} \boldsymbol{B}\cdot\boldsymbol{\delta B} \right),
\end{equation}
so in Fourier space
\begin{equation}
	\boldsymbol{\delta \tilde{\Pi}} = \delta \tilde{p} \mathsf{I} - \frac{1}{\mu_0}\left(\boldsymbol{B}\otimes \boldsymbol{\delta\tilde{B}} + \boldsymbol{\delta\tilde{B}}\otimes \boldsymbol{B} - \mathsf{I} \boldsymbol{B}\cdot\boldsymbol{\delta \tilde{B}} \right).
\end{equation}
Combining this with equation\ \eqref{eq:mhd} and the Boussinesq approximation (see Appendix\ \ref{appen:b}) we find
\begin{equation}
	\boldsymbol{q}\cdot\boldsymbol{\delta\tilde{\Pi}} = \boldsymbol{q} \delta p - \frac{i}{\mu_0} (\boldsymbol{B}\cdot\boldsymbol{q})^2 \boldsymbol{\delta\tilde{r}}.
\end{equation}
Note that as did\ \citet{2012MNRAS.426.1546B} we take $\boldsymbol{B}\cdot\boldsymbol{q}$ to be constant in time as implied by the Boussinesq and ideal-MHD conditions.
We now depart from prior work and use this equation along with equation\ \eqref{eq:momentum} taking the component perpendicular to $\boldsymbol{q}$ to eliminate $\delta p$ and find
\begin{align}
0 = &\left(\partial_t \delta\tilde{\boldsymbol{v}} + 2\boldsymbol{\Omega}\times\delta\tilde{\boldsymbol{v}}+ \hat{R} R \delta\tilde{\boldsymbol{r}}\cdot\nabla\Omega^2\right.\nonumber\\
 &- \left.\frac{1}{\gamma \rho}\left(\boldsymbol{\delta \tilde{r}}\cdot\nabla\sigma\right)\nabla\cdot\boldsymbol{\Pi} + \frac{1}{\mu_0 \rho} (\boldsymbol{B}\cdot\boldsymbol{q})^2 \boldsymbol{\delta \tilde{r}}\right)_{\perp \boldsymbol{q}},
\label{eq:momentum2}
\end{align}
where the notation $\left(...\right)_{\perp \boldsymbol{q}}$ denotes the component perpendicular to $\boldsymbol{q}$.

To construct the matrix version $\mathsf{L}$ of these equations we must choose a coordinate system.
Both because of the constraint\ \eqref{eq:bous} and because equation\ \eqref{eq:momentum2} is written in the plane perpendicular to $\boldsymbol{q}$ we choose the unit vectors
\begin{align}
	\hat{\boldsymbol{a}} &\equiv \frac{\hat{\boldsymbol{q}}\times\hat{\boldsymbol{w}}}{\sqrt{1-(\hat{\boldsymbol{q}}\cdot\hat{\boldsymbol{w}})^2}}\\
\intertext{and}
	\hat{\boldsymbol{b}} &\equiv \hat{\boldsymbol{q}}\times\hat{\boldsymbol{a}},
\end{align}
where $\hat{w}$ is any unit vector with $\hat{\boldsymbol{w}} \cdot \hat{\boldsymbol{q}} \neq 1$.
This choice of basis ensures that our vectors are perpendicular to the wavevector.

A choice of particular convenience for $\hat{w}$ is
\begin{equation}
	\hat{\boldsymbol{w}} = \frac{\nabla\Omega}{|\nabla\Omega|}.
\end{equation}
With this choice $\hat{a}$ is time-independent, because the component of $\boldsymbol{q}$ perpendicular to $\boldsymbol{w}$ is time-independent, and so we may write
\begin{equation}
	\boldsymbol{\delta\tilde{r}} = \alpha \hat{\boldsymbol{a}} + \beta \hat{\boldsymbol{b}}
\end{equation}
and
\begin{equation}
	\boldsymbol{\delta\tilde{v}} = \dot{\alpha} \hat{\boldsymbol{a}} + \dot{\beta} \hat{\boldsymbol{b}} + \beta \partial_t \hat{\boldsymbol{b}}.
\end{equation}
Note that there is a removeable singularity when $\hat{w} \parallel \hat{q}$.
The matrix $\mathsf{L}$ is then given by computing the relation between $\partial_t \left\{\alpha,\beta,\dot{\alpha},\dot{\beta}\right\}$ and $\left\{\alpha,\beta,\dot{\alpha},\dot{\beta}\right\}$.
The result is quite unwieldy so we do not present it here but note that it is fully documented in the software in which we implement these equations.

\section{Stresses and Transport}
\label{sec:stress}

The equations of motion contain the position and the velocity, so our expanded vector space is
\begin{equation}
	\Phi = \left\{\delta\boldsymbol{r}, \delta\boldsymbol{v}, \partial_t \delta \boldsymbol{v}, ..., \right\}.
\end{equation}
Combining the linearised equations of motion with our closure scheme we can compute the correlation function
\begin{equation}
	\langle \Phi \otimes \Phi \rangle = \int \frac{d^3 \boldsymbol{k}}{(2\pi)^3} \sum_i \langle\Phi^{i}_{\boldsymbol{k}}\otimes\Phi^{i*}_{\boldsymbol{k}}\rangle,
\end{equation}
where the index $i$ ranges over eigenvectors.
This function contains all of the usual stresses and transport functions.
For instance, the Reynolds stress is
\begin{equation}
	R \equiv \langle \delta\boldsymbol{v}\otimes\delta\boldsymbol{v}\rangle = \langle \Phi_1 \otimes \Phi_1 \rangle.
\end{equation}
Likewise up to a dimensionless constant of order unity the turbulent diffusivity is
\begin{equation}
	d \equiv \langle \delta\boldsymbol{v}\otimes\delta\boldsymbol{r}\rangle = \langle \Phi_1 \otimes \Phi_0 \rangle.
\end{equation}
and the turbulent viscosity is
\begin{equation}
	Q \equiv \langle \delta\boldsymbol{v}\otimes\delta\boldsymbol{r}\rangle+\langle \delta\boldsymbol{r}\otimes\delta\boldsymbol{v}\rangle = \langle \Phi_1 \otimes \Phi_0 \rangle+\langle \Phi_0 \otimes \Phi_1 \rangle.
\end{equation}
Similar expressions hold for the dynamo effect, the transport of magnetic fields, and material diffusion.

\section{Results}
\label{sec:results}

In this section we exhibit a number of results which come from applying our model to a wide variety of astronomically- and physically-relevant circumstances.
We also compare with the results of\ \citet{2013MNRAS.431.2200L} and\ \citet{1993A&A...276...96K}.
We modify the former to use the convention in equation\ \eqref{eq:k0} to avoid spurious differences in scale.
We likewise assume that our $L_0$ is equal to three times the mixing length of\ \citet{1993A&A...276...96K}, as this is an inherent freedom in the formalism and resolves an otherwise-persistent scale difference between our model and theirs.
These models have been well-tested against a variety of data, most notably helioseismic results, and so provide a useful reference for our work.

We have also included more direct comparisons but, because direct experiments are extremely difficult to perform under most circumstances relevant to astrophysics, we have instead included comparisons with simulations and observations where available and applicable.
Simulations are often the most useful comparison for stellar phenomena, because a variety of processes, including meridional circulation, can mask the effects of turbulent transport~\citep{2013IAUS..294..399K}.
In accretion discs, however, there are several observable quantities which are thought to correlate closely with the underlying turbulence and these provide very helpful constraints~\citep{2007MNRAS.376.1740K}.

These comparisons and calculations are not intended to be a complete collection of the results our model can produce, nor have we exhaustively explored the circumstances and dependencies of each result.
Rather it is our hope to demonstrate that there is a great deal of interesting physics in this model, that our perturbative corrections give rise to realistic results and reproduce many known results, and that there is much to warrant further exploration along these lines.

\subsection{Rotating Convection}

\begin{figure*}
	\centering
	\includegraphics[width=0.95\textwidth]{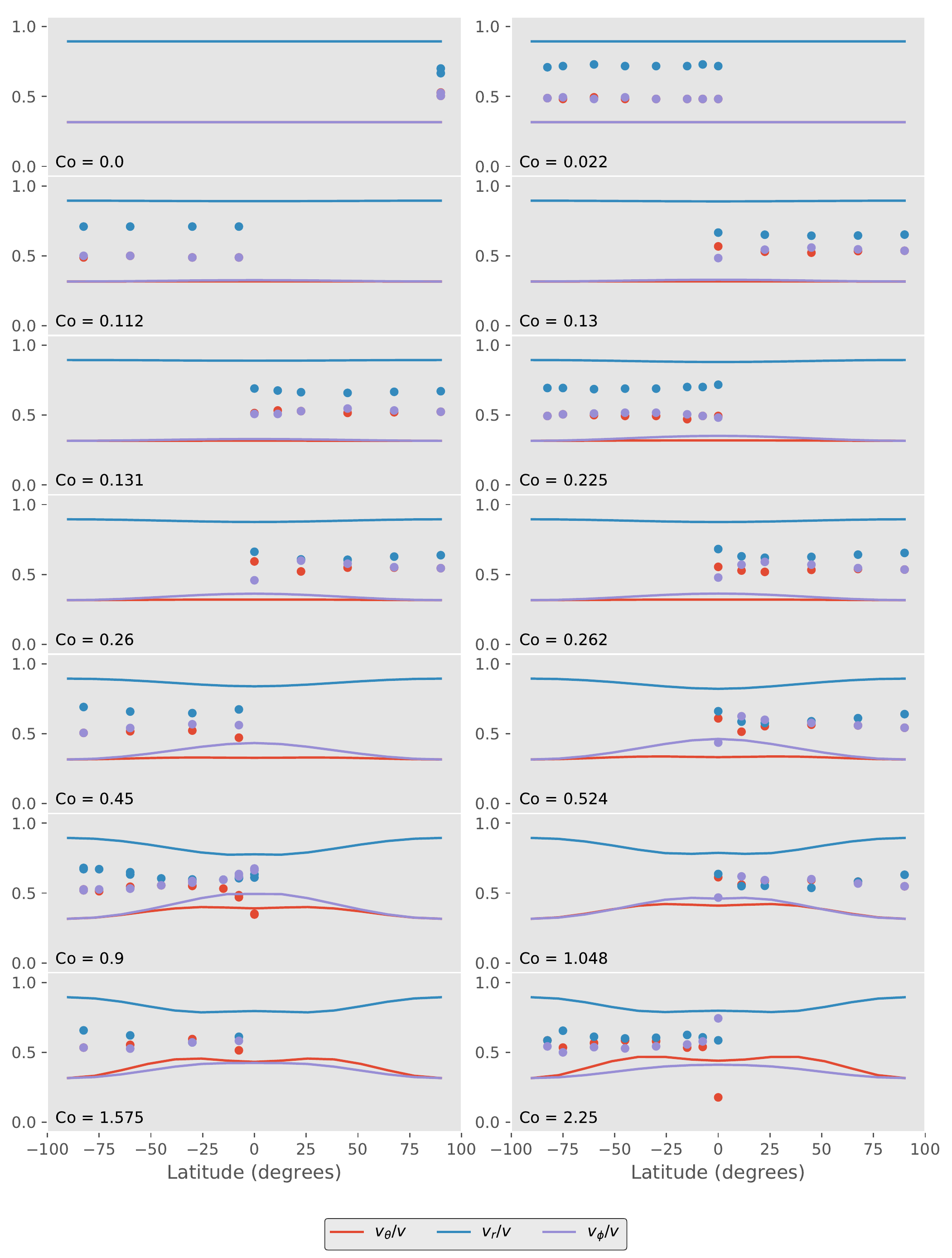}
	\caption{The ratios $\sqrt{\langle \delta v_r^2\rangle/\langle \delta v^2\rangle}$ (blue), $\sqrt{\langle \delta v_\theta^2\rangle/\langle \delta v^2\rangle}$ (red) and $\sqrt{\langle \delta v_\phi^2\rangle/\langle \delta v^2\rangle}$ (purple) are shown for our model (solid) and for simulations by~\citep[][dots, negative latitude]{2004A&A...422..793K} and~\citep[][dots, positive latitude]{2001ApJ...548.1102C} for a wide range of rotation rates as a function of latitude. The rotation rate is captured by the Coriolis number $\mathrm{Co}=\Omega h / \langle \delta v^2\rangle^{1/2}$. Our model general overestimates the anisotropy but captures its variation well.}
	\label{fig:kaplya1}
\end{figure*}

We begin with the effect of rotation on convection in the case of a rotating system with radial pressure and entropy gradients.
It is useful to start by comparing our results with those from simulations.
Fig.~\ref{fig:kaplya1} shows the ratios $\sqrt{\langle \delta v_r^2\rangle/\langle \delta v^2\rangle}$, $\sqrt{\langle \delta v_\theta^2\rangle/\langle \delta v^2\rangle}$ and $\sqrt{\langle \delta v_\phi^2\rangle/\langle \delta v^2\rangle}$ for several rotation rates as a function of latitude.
The positive latitudes come from Table 2 of ~\citet{2001ApJ...548.1102C} while the negative are from Table 2 of~\citet{2004A&A...422..793K}.
In order to match the units for the rotation rates we put everything in terms of the coriolis number
\begin{align}
	\mathrm{Co} \equiv \frac{\Omega h}{\langle \delta v^2\rangle^{1/2}},
\end{align}
where, following the convention of ~\citet{2004A&A...422..793K}, $\langle \delta v^2\rangle^{1/2}$ was computed for a non-rotating system.

Our model overestimates the anisotropy of the turbulence but captures its symmetries and trends well.
For instance we find that near the poles and in non-rotating systems the $\theta$ and $\phi$ components of the velocity fluctuations have identical magnitudes, in line with the simulations.
We reproduce the trend of decreasing anisotropy towards the equator and decreasing anisotropy with increasing rotation, and, in cases where there are differences between the $\theta$ and $\phi$ velocities, we reproduce both their sign and magnitude.
In particular we find that $\langle \delta v_r^2\rangle \geq \langle \delta v_\theta^2 \rangle \geq \langle \delta v_\phi^2\rangle$, which is seen in these and other simulations~\citep{2005AN....326..315R}.
Likewise we find that radial motion makes up a greater fraction of the total velocity near the poles than at the equator, and that as the Coriolis number increases $\langle \delta v_r^2 - \delta v_\theta^2 - \delta v_\phi^2 \rangle \rightarrow 0$, all of which is in agreement with the predictions of~\citet{2005A&A...431..345R}.

Our overestimate of the anisotropy may be due to our model incorporating the large-scale fields on all scales, as noted by~\citet{2013MNRAS.431.2200L}.
This suggests that a future refinement might be to use estimates of the large-scale modes to compute the environment of those at smaller scales, but we do not treat such complications for now.

\begin{figure}
	\centering
	\includegraphics[width=\figwidth]{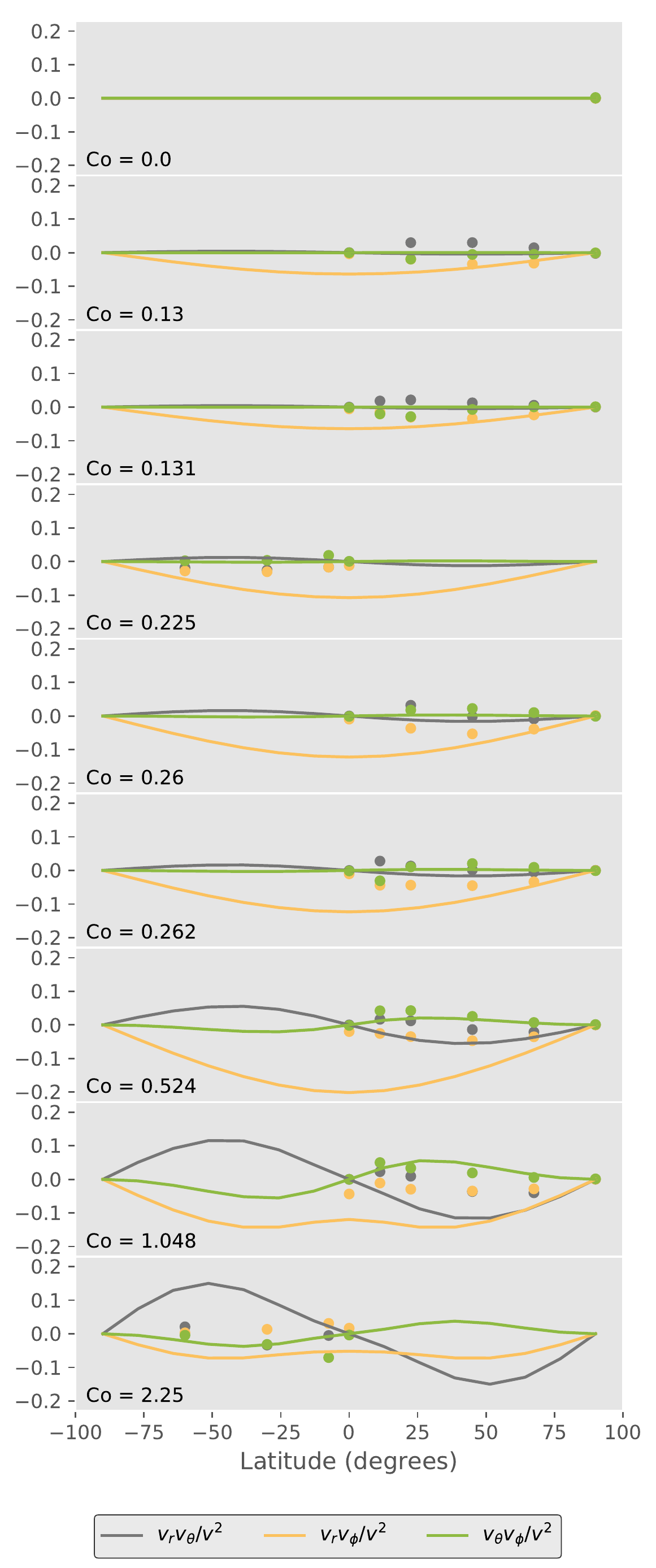}
	\caption{The ratios $\langle \delta v_r \delta v_\theta \rangle/\langle \delta v^2\rangle$ (red), $\langle \delta v_\theta \delta v_\phi \rangle/\langle \delta v^2\rangle$ (purple) and $\langle \delta v_r \delta v_\phi\rangle/\langle \delta v^2\rangle$ (blue) are shown from our model (solid) and from simulations by~\citep[][dots, negative latitude]{2004A&A...422..793K} and~\citep[][dots, positive latitude]{2001ApJ...548.1102C} as a function of latitude. Note that~\citet{2004A&A...422..793K} cautions that the moderate rotation simulations had difficulty converging, and these results arise as the difference between two simulations, so it is not clear how significant this test is. Our model generally overestimates these stresses, and suggests a different symmetry for the variation (going as $\sin\theta$ rather than $\sin(2\theta)$).}
	\label{fig:kaplya3}
\end{figure}

As a further comparison we consider the off-diagonal Reynolds stresses of both~\citet{2001ApJ...548.1102C} and~\citet{2004A&A...422..793K}.
These numbers were extracted from Table~3 of the former and also Table~3 of the latter and are shown along with our predictions in Fig.~\ref{fig:kaplya3}.
In the former they were straightforward to analyse but in the latter they do not provide a precise test because the simulations included a bulk shear.
To correct for this we used a linear expansion to subtract results across simulations which were identical in all conditions other than the rotation and thereby determine the effect of the rotation alone.
As we will see in Section~\ref{sec:diffrotconv} this procedure is problematic because the shear may interact non-linearly with the rotation.
Furthermore because these corrections are of the same order as the terms themselves some care must be taken in interpreting the results.

Despite these difficulties some trends are clear and sustained between both sets of data.
For instance in the northern hemisphere ($\theta > 0$), $\langle \delta v_r \delta v_\theta \rangle < 0$, while in both hemispheres $\langle \delta v_r \delta v_\phi\rangle < 0$, in keeping with predictions and simulations by~\cite{2005A&A...431..345R}.
Likewise we find that $\langle v_\theta v_\phi \rangle > 0$ in the northern hemisphere, in agreement with the findings of~\citet{2005AN....326..315R}.

Once more, however, our model overestimates these anisotropic terms by an amount which is largely invariant as a function of rotation.
This suggests that this overestimate is a systematic offset rather than an error in scaling.
We also have some difficulty to reproduce the signs of some of the stresses, particularly in the results of~\citet{2004A&A...422..793K}, though this could simply be a subtraction difficulty.
This is supported by the fact that the simulations themselves do not agree on the signs of these terms and highlights the challenges of making comparisons of terms which are small in magnitude relative to the scale of the turbulence.

\begin{figure}
	\centering
	\includegraphics[width=\figwidth]{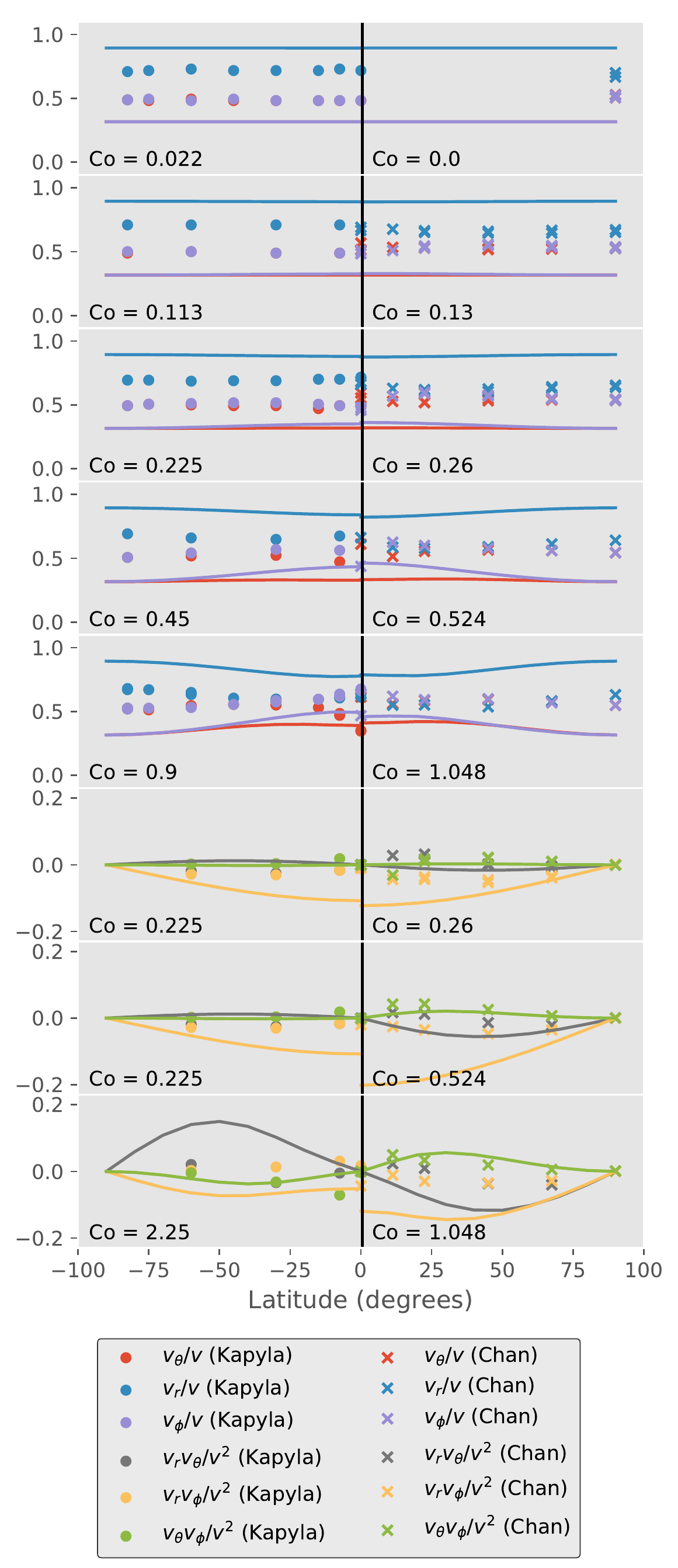}
	\caption{The functions shown in Figs.~\ref{fig:kaplya1} and~\ref{fig:kaplya3} are shown from our model (solid), simulations by~\citep[][dots, negative latitude]{2004A&A...422..793K} and~\citet{2001ApJ...548.1102C} (crosses, positive latitude) as a function of latitude. The most comparable pairs of rotation rates were placed side-by-side for each function. A solid black line is shown along the equator where the latitude is zero. There is reasonable agreement on the distribution of velocities in direction but not on the correlations between different velocity directions..}
	\label{fig:comparison}
\end{figure}

To better understand which trends are significant and which are artefacts we have placed data from comparable rotation rates for the two sets of simulations side-by-side in Fig.~\ref{fig:comparison}.
The top five panels show the same data as in Fig.~\ref{fig:kaplya1} while the bottom three show the data from Fig.~\ref{fig:kaplya3}.
In general there is good agreement in the top five panels.
The data of~\citet{2004A&A...422..793K} gives systematically larger anisotropies and the two sets of simulations occasionally differ on the relative magnitudes of the velocity components (i.e. their ordering), but otherwise the two are in good agreement.
By contrast the bottom three panels paint two very divergent pictures.
Neither ordering, trends nor signs are consistent between the two sets of simulations.
Only the magnitudes agree in these cases.
Thus the two sets of simulations agree that our model systematically overestimates anisotropies and that, beyond that, our model agrees with them to the extent that they agree with one another.

\begin{figure}
	\centering
	\includegraphics[width=\figwidth]{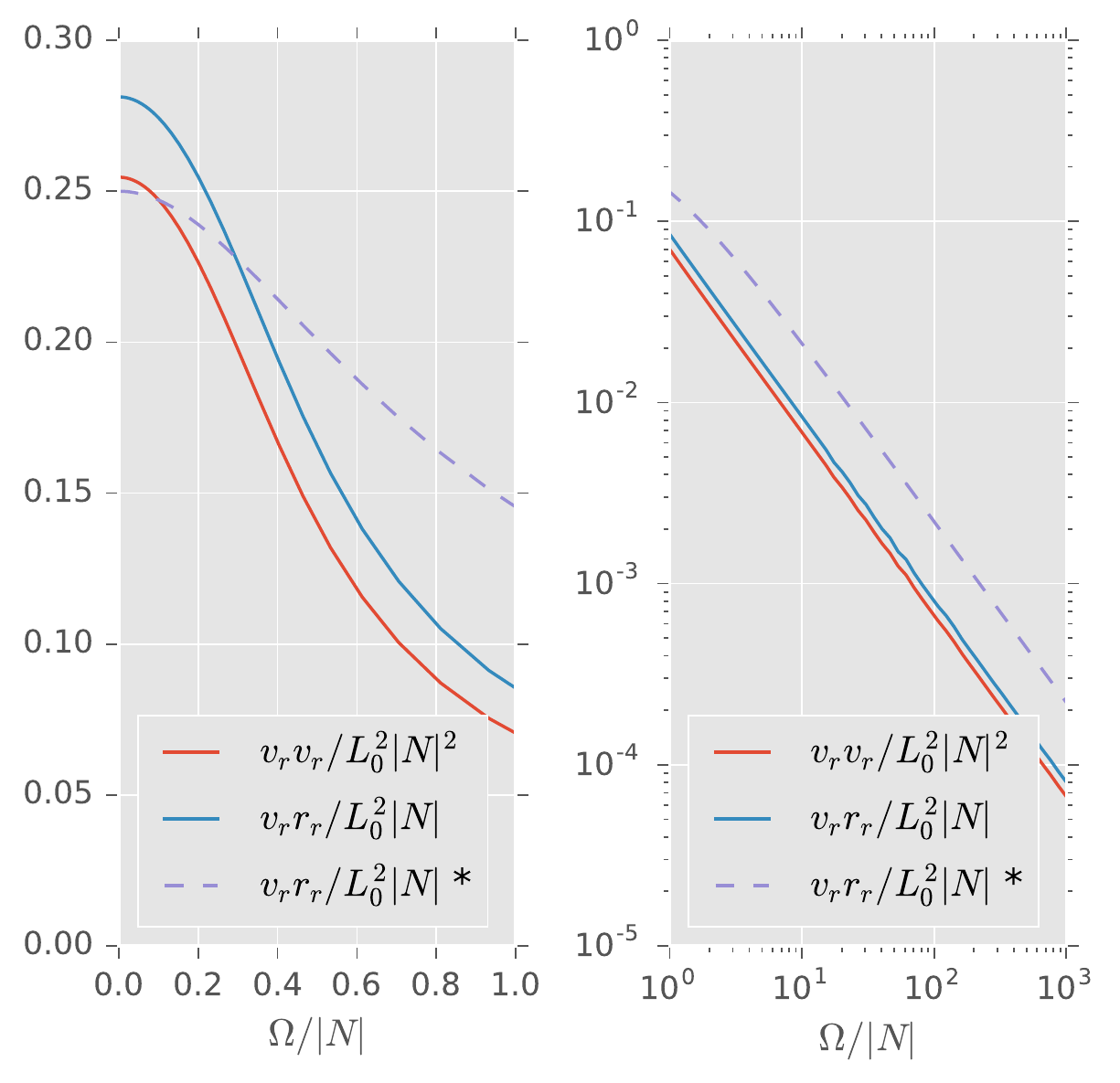}
	\caption{The radial velocity correlation function $\langle \delta v_r \delta v_r \rangle$ (red) and the radial diffusivity $\langle \delta v_r \delta r_r \rangle$ (blue) are shown in linear scale for $\Omega < |N|$ (left) and log-log scale for $\Omega > |N|$ (right). These results are for uniform rotation at a latitude of $\pi/4$ with no magnetic field.  On this and all subsequent figures $v_r v_r/L_0^2|N|^2$ should be read as $\langle \delta v_r \delta v_r \rangle / L_0^2|N|^2$ and similarly for other correlations. Shown in purple (*, dashed) for comparison is the result of \citet{1993A&A...276...96K} with an anisotropy factor of $2$, which agrees in sign, scale and variation. The bumps in our results reflect parameter values where the numerical integration was more difficult. All quantities are given in units of the mixing length and the \brvs\ frequency.}
	\label{fig:rotconv}
\end{figure}

Having compared in detail with these simulations we now consider predictions which go beyond the domain where simulations are possible.
In convection with radial gradients the leading order effect is to transport heat and material radially.
Fig.\ \ref{fig:rotconv} shows $\langle \delta v_r \delta v_r \rangle$ and $\langle \delta v_r \delta r_r \rangle$, which are the correlation functions controlling this transport.

Both correlators vary at second order in $\Omega$ in the slow rotation limit as expected\ \citep{2013MNRAS.431.2200L, 2013IAUS..294..399K}.
In the rapid rotation limit on the other hand they exhibit clear $\Omega^{-1}$ scaling, consistent with what is seen in other closure models and in simulations\ \citep{2010MNRAS.407.2451G}.
The quenching of turbulence in this limit arises because the Coriolis effect acts as a restoring force, stabilising modes.

The peak of each correlator is of order unity and occurs when $\Omega = 0$.
In fact for the stress the maximum is $0.254647$ while for the diffusivity it is $0.28125$, both of which are consistent to this precision with \citet{2013MNRAS.431.2200L}, noting that we used the definition in equation\ \eqref{eq:k0} for their mixing length.
This is because our model is precisely the same as theirs in this limit.
Based on this and the observed scalings a good approximation is
\begin{align}
	\langle \delta v_r \delta r_r \rangle \approx \langle \delta v_r \delta v_r \rangle \approx \frac{1-(\Omega/|N|)^2}{1 - (\Omega/|N|)^3}.
\end{align}

\begin{figure}
	\centering
	\includegraphics[width=\figwidth]{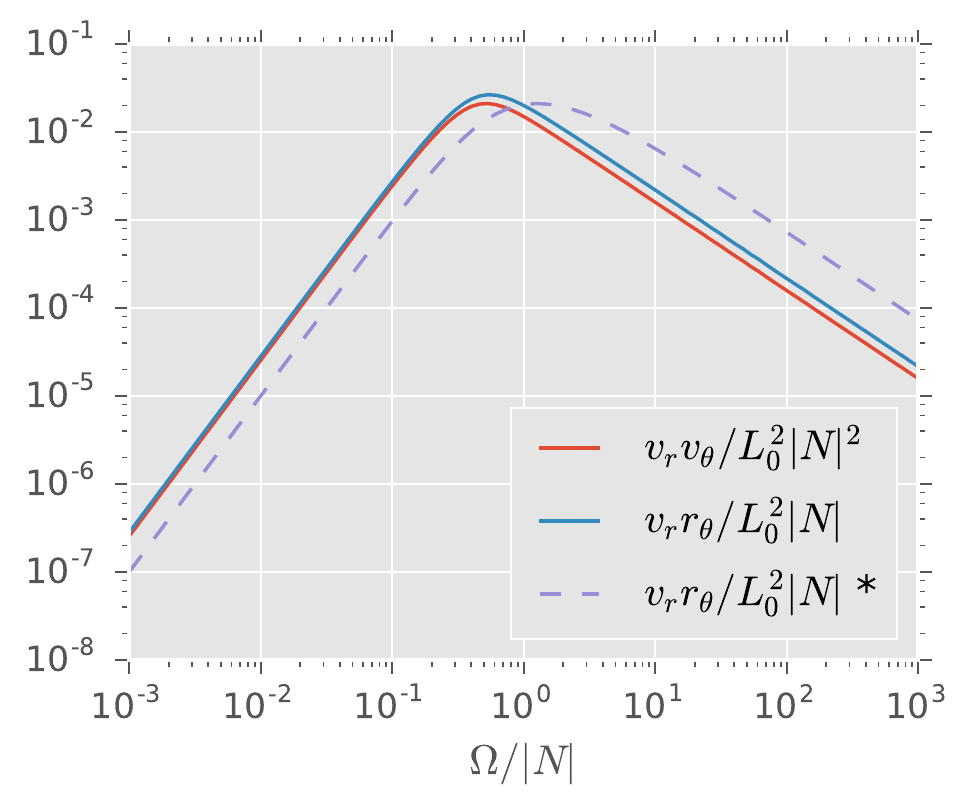}
	\caption{The absolute value of the $r-\theta$ velocity correlation function $\langle \delta v_r \delta v_\theta \rangle$ (red) and corresponding diffusivity $\langle \delta v_r \delta r_r \rangle$ (blue) are shown in log-log scale against rotation rate. These results are for uniform rotation at a latitude of $\pi/4$ with no magnetic field. Shown in purple (*, dashed) for comparison is the result of \citet{1993A&A...276...96K} with an anisotropy factor of $2$, which agrees in sign, scale and variation. All quantities are given in units of the mixing length and the \brvs\ frequency.}
	\label{fig:rotconv2}
\end{figure}

Next we consider the effect of rotation on the $r-\theta$ correlation functions.
These functions are responsible for latitudinal transport of heat, mass and momentum and vanish as a result of spherical symmetry in the non-rotating limit.
Fig.\ \ref{fig:rotconv2} shows $\langle \delta v_r \delta v_\theta \rangle$ and $\langle \delta v_r \delta r_\theta \rangle$ as a function of the rotation rate.

In the slow-rotation regime both quantities scale as $\Omega^2$, while in the rapid rotation limit they scale as $\Omega^{-1}$.
The peak is of order unity and occurs near $\Omega = |N|$.
This gives rise to the approximation
\begin{align}
	\langle v_r r_\theta \rangle \approx \langle v_r v_\theta \rangle \approx \frac{(\Omega/|N|)^2}{1 + (\Omega/|N|)^3}.
\end{align}
These scalings may be interpreted as a competition between symmetry breaking and quenching: the correlation function rises as rotation breaks symmetries but excessive rotation stabilises the system and quenches the turbulent motions.
The symmetry is broken quadratically because, at first order, the Coriolis effect only couples radial and azimuthal motions.

\begin{figure}
	\centering
	\includegraphics[width=\figwidth]{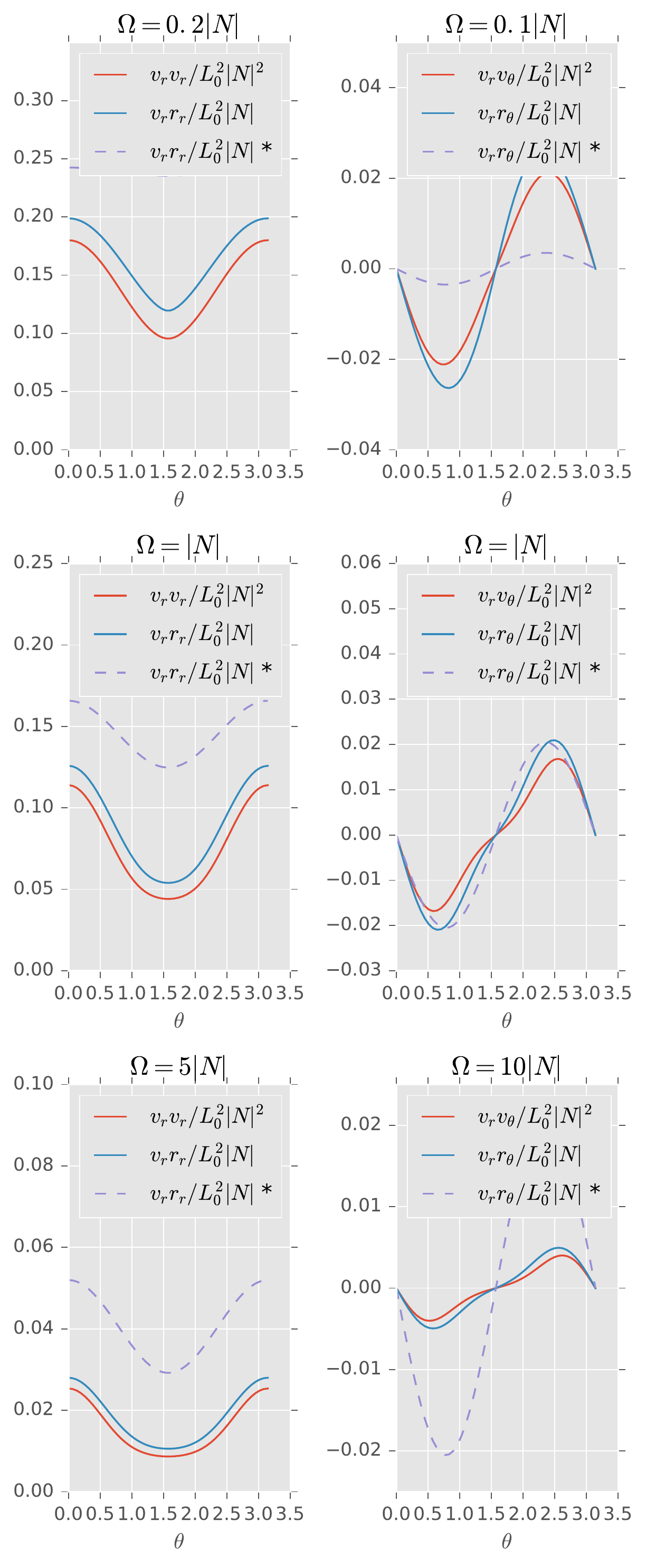}
	\caption{Various correlation functions are shown as a function of the angle $\theta$ from the rotation axis. The functions are the $r-r$ (left) and $r-\theta$ (right) velocity (red) and diffusivity (blue) correlation functions. These results are for uniform rotation at $\Omega=0.2|N|$ (top), $\Omega=|N|$ (middle) and $\Omega=5|N|$ (bottom). Shown in purple (*, dashed) for comparison is the KR result, which agrees in sign and variation but not scale. For slow rotation the scale of the variation is generally smaller than we predict, while for fast rotation the variation is somewhat larger. All quantities are given in units of the mixing length and the \brvs\ frequency.}
	\label{fig:rotconv4}
\end{figure}

The properties of turbulence vary with latitude in a rotating system because the rotation axis picks out a preferred direction.
Fig.~\ref{fig:rotconv4} shows the $r-r$ and $r-\theta$ stress and diffusivity correlations as a function of latitude.
The $r-r$ correlations vary similarly to one another, exhibiting a minimum at the equator and maxima on-axis.
On-axis the rotation drops out of the equations and so the on-axis functions are just those for non-rotating convection.
The effect of rotation is then largest at the equator, where the convective motion is predominantly perpendicular to the rotation axis.
The correlation functions are smallest where the rotation has the largest effect because rotation primarily acts to stabilise modes.

By contrast the $r-\theta$ correlator is largest in magnitude at mid-latitudes, vanishing both on-axis and at the equator.
On-axis this correlation function must vanish because the $\hat{\theta}$ unit vector is ill-defined.
The sign change between the northern and southern hemispheres occurs because $(\hat{\boldsymbol{r}} \times \boldsymbol{\Omega})_\phi$ has the same sign everywhere while $(\hat{\boldsymbol{\theta}} \times \boldsymbol{\Omega})_\phi$ changes sign between the hemispheres.
This also explains the vanishing correlation at the equator.

\begin{figure}
	\centering
	\includegraphics[width=\figwidth]{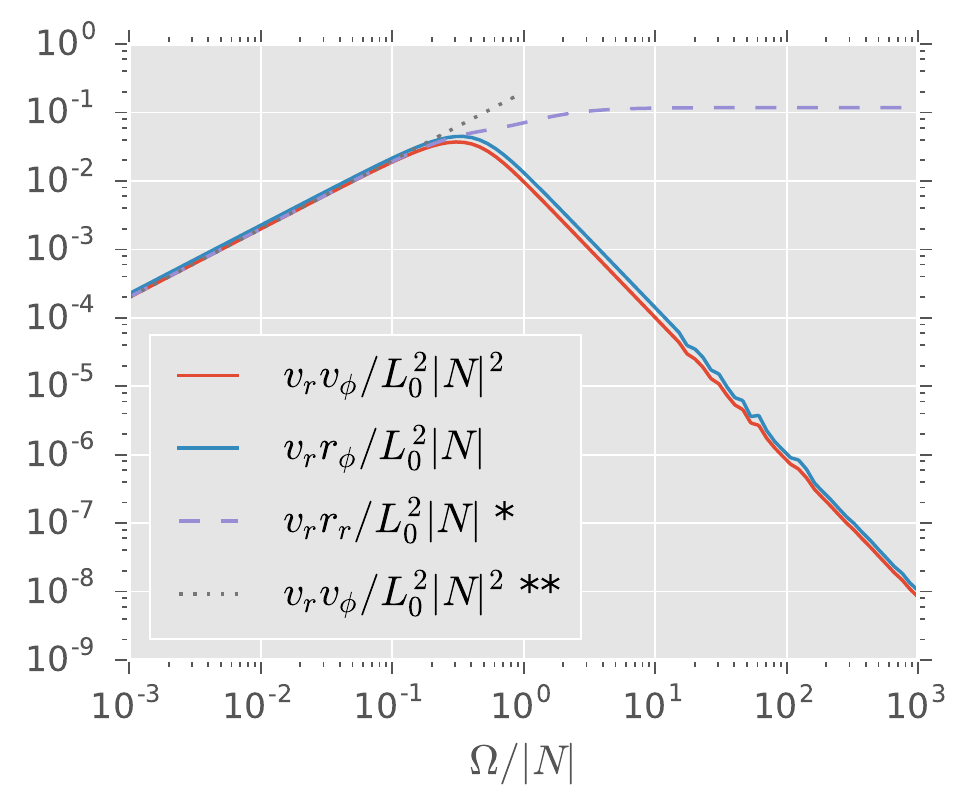}
	\caption{The absolute value of the $r-\phi$ velocity correlation function $\langle \delta v_r \delta v_\phi \rangle$ (red) and corresponding diffusivity $\langle \delta v_r \delta r_\phi \rangle$ (blue) are shown in log-log scale versus rotation rate. These results are for uniform rotation at a latitude of $\pi/4$ with no magnetic field.  Shown in purple (*, dashed) for comparison is the result of \citet{1993A&A...276...96K} with an anisotropy factor of $2$ which agrees in sign, variation and scale up until $\Omega=|N|$, at which point the behaviour differs significantly. Shown in grey (**, dotted) for comparison is $\langle \delta v_r \delta v_\phi \rangle$ from that of \citet{2013MNRAS.431.2200L}. This agrees precisely in the $\Omega \rightarrow 0$ limit and the agreement is good even near $\Omega \approx 0.5|N|$. All quantities are given in units of the mixing length and the \brvs\ frequency.}
	\label{fig:rotconv3}
\end{figure}

The quantities of particular interest for studying the origins of differential rotation are the radial-azimuthal correlation functions $\langle \delta v_r \delta v_\phi \rangle$ and $\langle \delta v_r \delta r_\phi \rangle$.
The former provides a stress coupling the angular momentum to radial motions known as the $\Lambda$-effect, while the latter provides a viscosity coupling radial shears to azimuthal motion and so acts as a proxy for the $\alpha$-effect\ \citep{1993A&A...276...96K}.
Fig.\ \ref{fig:rotconv3} shows these quantities as a function of the rotation rate.
In the slow-rotation limit both scale as $\Omega$ before peaking near unity and falling off as $\Omega^{-2}$ in the rapid-rotation limit.
The linear scaling at slow rotation rates is a consequence of the Coriolis effect directly coupling radial and azimuthal motions.
These quantities fall off more rapidly than the others in the case of rapid rotation because it is preferentially the modes which couple strongly to the Coriolis effect which are stabilised the most.
The absolute scale of our $\Lambda$-effect is approximately what is seen in simulations, slightly overestimating relative to~\citet{2004A&A...422..793K} and similar to other theoretical predictions~\citep{2013IAUS..294..399K,2012ISRAA2012E...2G}.

\subsection{Differential Rotation and Convection}
\label{sec:diffrotconv}

We now turn to the dependence of convective transport coefficients on differential rotation.
We expand our closure model to linear order in the shear and so restrict this analysis to cases where the dimensionless shear $|R\nabla\ln\Omega|$ is at most of order unity.

\begin{figure}
	\centering
	\includegraphics[width=\figwidth]{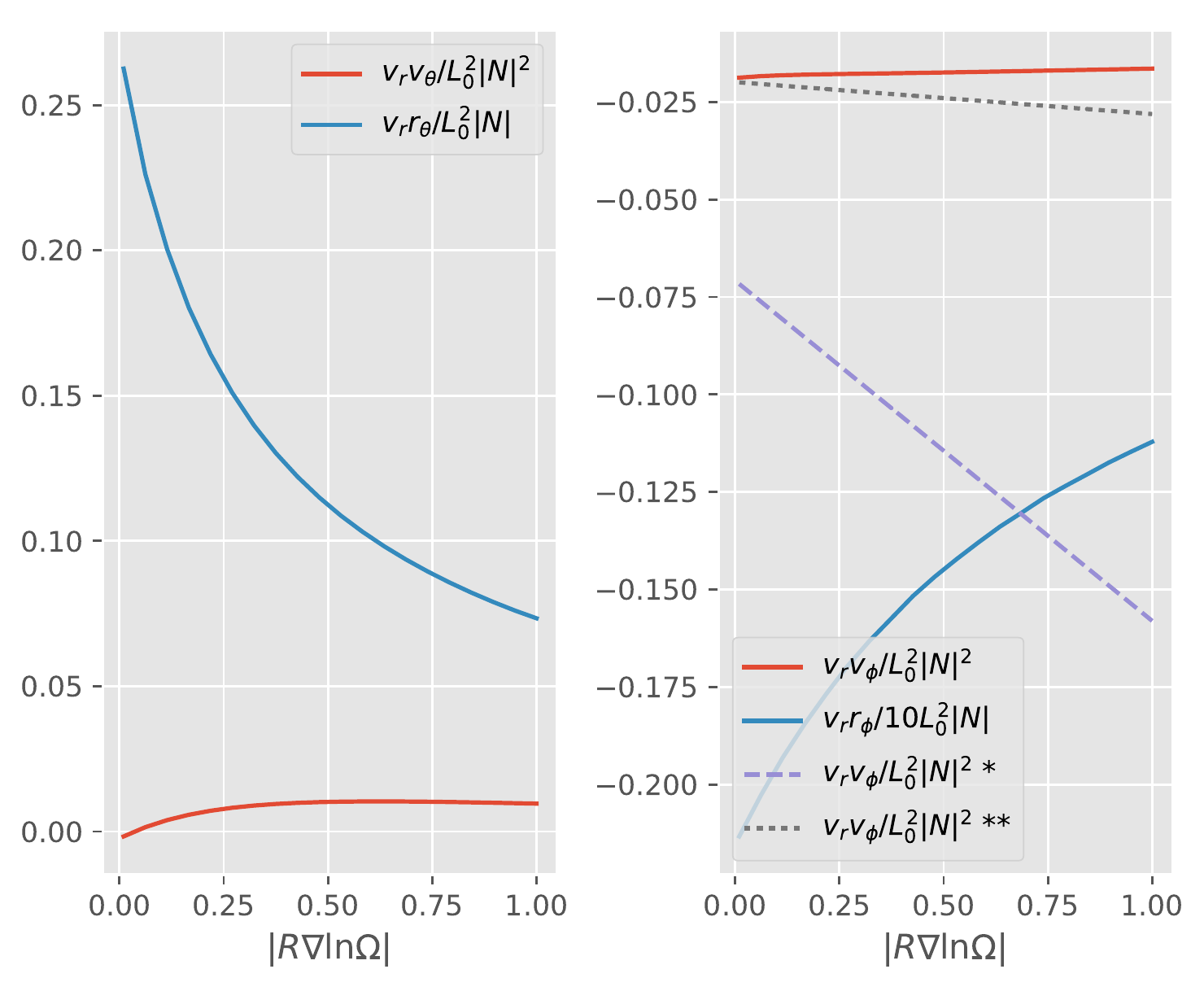}
	\caption{The $r-\theta$ (left) and $r-\phi$ (right) velocity (red) and diffusivity (blue) correlation functions are shown in log-scale versus the differential rotation. These results are for a convecting region with differential rotation in the cylindrical radial direction, $\Omega = 0.1 |N|$ and no magnetic field at a latitude of $\pi/4$. Shown in purple (*, dashed) for comparison is the result of \citet{1993A&A...276...96K} with an anisotropy factor of $2$. This disagrees on the magnitude of the slope but agrees in the sign of the slope. Shown in grey (**, dotted) for comparison is $\langle \delta v_r \delta v_\phi \rangle$ of \citet{2013MNRAS.431.2200L}. This generally predicts smaller stresses though with the same sign and slope sign as our model. All quantities are given in units of the mixing length and the \brvs\ frequency.}
	\label{fig:diffrot}
\end{figure}

Fig.\ \ref{fig:diffrot} shows the $r-\theta$ and $r-\phi$ velocity and diffusivity correlation functions as a function of differential rotation for a situation where $\nabla\Omega$ is at an angle of $\pi/4$ relative to the pressure gradient.
All four functions behave linearly near the origin, with intercept set by the stress and diffusivity in the uniform rotation limit.
This is precisely as expected: the intercept is non-zero, giving rise to the $\Lambda$-effect, while the slope is non-zero, giving rise to the $\alpha$-effect\ \citep{1993A&A...276...96K}.
Note that the favourable comparison of our results with those of~\citet{2013IAUS..294..399K} are helpful because their model was implemented in a two-dimensional solar model which compared well with helioseismic observations.

A key difference between our work and what we compare with in Fig.\ \ref{fig:diffrot} is that, while we predict the same sign and comparable magnitude for the $\alpha$-effect in the zero-shear limit, the effect greatly reduces near $R\nabla\ln\Omega|\approx 0.1$, indicating that, at least for this configuration, this is the point at which non-linear effects become important.
This does not represent a particularly severe shear and highlights a key point that the correlation functions we find are generally non-linear in all of the small parameters in which one might wish to expand.
Our model captures this nonlinear behaviour despite being carried out to linear order in $|R\nabla\ln\Omega|$.
This is because, in our expansion, the time evolution operator is what is expanded linearly.
The resulting eigenvalues and eigenvectors are generally non-linear functions of this operator.

\begin{figure}
	\centering
	\includegraphics[width=\figwidth]{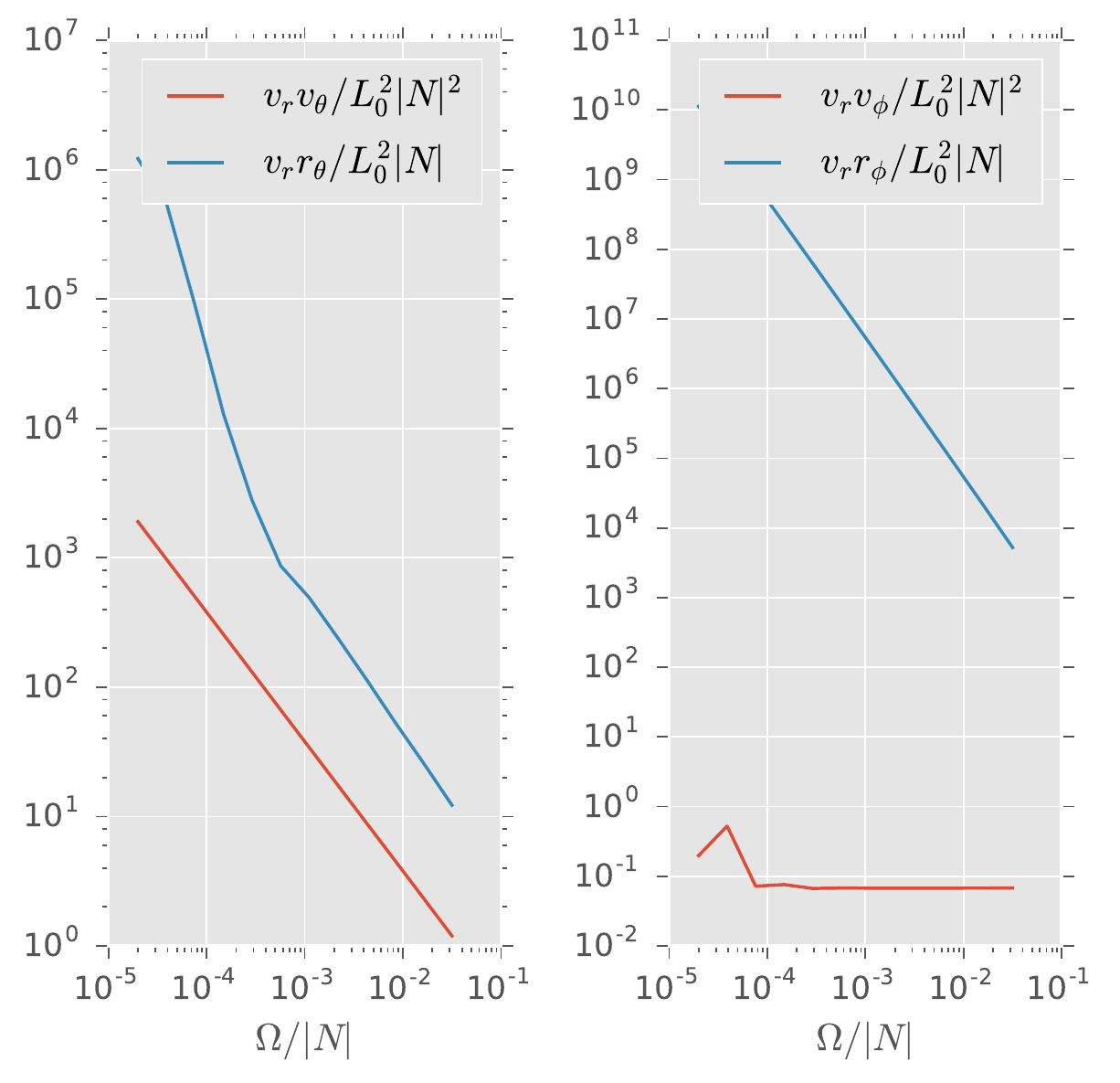}
	\caption{The derivatives of various correlation functions with respect to $|R\nabla\Omega|$ are shown as a function of $\Omega$, with both axes log-scaled. The functions are the $r-\theta$ (left) and $r-\phi$ (right) velocity (red) and diffusivity (blue) correlation functions. These results are for a convecting region with differential rotation in the cylindrical radial direction and no magnetic field at a latitude of $\pi/4$. All quantities are given in units of the mixing length and the \brvs\ frequency.}
	\label{fig:diffrot2}
\end{figure}

This caution aside, there is a significant regime where the $\alpha-\Lambda$ expansion is valid and, in this regime, key quantities of interest are the derivatives of the various correlation functions with respect to the shear $|R\nabla\Omega|$.
Fig.\ \ref{fig:diffrot2} shows these derivatives as a function of $\Omega$.
The $r-\phi$ stress derivative is constant in $\Omega$.
This means that the stress scales as $R\nabla\Omega$.
This is as expected~\citep[see, e.g. Equation 79 of][]{2013MNRAS.431.2200L} and indicates that there is a well-defined effective viscosity transporting angular momentum.
This viscosity is given by
\begin{align}
	\nu_{r\phi} \approx L_0^2 |N|.
	\label{eq:nurphi}
\end{align}

By contrast the derivatives of the $r-\theta$ correlations as well as the $r-\phi$ diffusivity all diverge in the limit as $\Omega \rightarrow 0$.
In particular, the $r-\theta$ velocity correlation diverges as $\Omega^{-1}$, the $r-\theta$ diffusivity correlation diverges as $\Omega^{-2}$ and the $r-\phi$ diffusivity diverges as $\Omega^{-2}$.
These divergences are signatures of symmetry breaking.
They indicate that the direction in which the $R\nabla \Omega \rightarrow 0$ limit is approached matters.
That is, this limit can be approached by first letting $\Omega \rightarrow 0$ and then differentiating or by differentiating and then taking $\Omega \rightarrow 0$ and the divergence we find in the latter approach indicates that the order matters.

When $\Omega=0$ and $|R\nabla\Omega|=0$ there is a symmetry between $\pm \theta$ and between $\pm \phi$.
As a result both the $r-\theta$ and $r-\phi$ terms vanish in this limit.
When $\Omega \neq 0$ these symmetries are broken by the rotation and we know from Figs.\ \ref{fig:rotconv2} and\ \ref{fig:rotconv3} that this occurs at first order for $r-\phi$ and second order for $r-\theta$.
In the opposing limit the situation is different because in the time evolution described by equation\ \eqref{eq:momentum2} $\mathsf{L}$ is independent of $|R\nabla\Omega|$ when $\Omega=0$.
There is, however, a dependence on $|R\nabla\Omega|$ through the time-dependence of $\boldsymbol{q}$.
This breaks the $\phi$ symmetry because $\partial_t \boldsymbol{q}$ is proportional to $q_\phi R\nabla\Omega$ and hence is sensitive to $\phi$.
It does not, however, break the $\theta$ symmetry, because $q_\phi R\nabla\Omega$ is symmetric with respect to changing the signs of both $\theta$ and $\boldsymbol{q}$. 
It follows then that we should find divergences in the $r-\theta$ correlation derivatives owing to the path-dependence of the zero-rotation limit and that we should find the $r-\phi$ derivatives to be generally well-behaved.

The curious divergence is then that in the $r-\phi$ diffusivity, because this correlation function does not suffer from a symmetry-derived path-dependence.
This arises because the differential rotation means that $\mathsf{L}$ is time-dependent.
This introduces polynomial corrections to the usual exponential growth, as discussed in Section\ \ref{sec:perturb}.
This formalism captures the fact that the differential rotation turns vertical displacement into $\phi$ displacements which vary as polynomials in time.
There are therefore modes with very small radial velocities which nevertheless have large azimuthal displacements and these dominate the diffusivity derivative.
These modes grow proportional to $|R\nabla\Omega|$ and their growth may proceed in the azimuthal direction until bounded by the Coriolis effect at a time $\Omega^{-1}$.
As a result these modes contribute to the diffusivity as $|R\nabla\ln\Omega|$ and hence lead to a diverging derivative in $|R\nabla\Omega|$ as $\Omega\rightarrow 0$.

\subsection{Differential Rotation and Stable Stratification}

Stably stratified regions are those with
\begin{align}
	N^2 > 0,
\end{align}
such that buoyancy acts to counter perturbations in the vertical direction.
This tends to damp turbulence.

In the presence of such damping there can still be turbulence if there is also a shear.
The classic example of this is the Kelvin-Helmholtz phenomenon, which can occur in such a system if the Richardson criterion
\begin{align}
	\frac{|du/dz|^2}{|N|^2} > \frac{1}{4}
\end{align}
is satisfied\ \citep{1993SSRv...66..285Z}.
Here $u$ is the velocity and $z$ is the coordinate parallel to the stratification.
Even when this criterion is not satisfied, latitudinal shear can still generate turbulence~\citep{2008JAtS...65.2437C}.
These motions are suppressed in vertical extent by the stratification and hence are primarily confined to the plane perpendicular to the stratification direction.

\begin{figure}
	\centering
	\includegraphics[width=\figwidth]{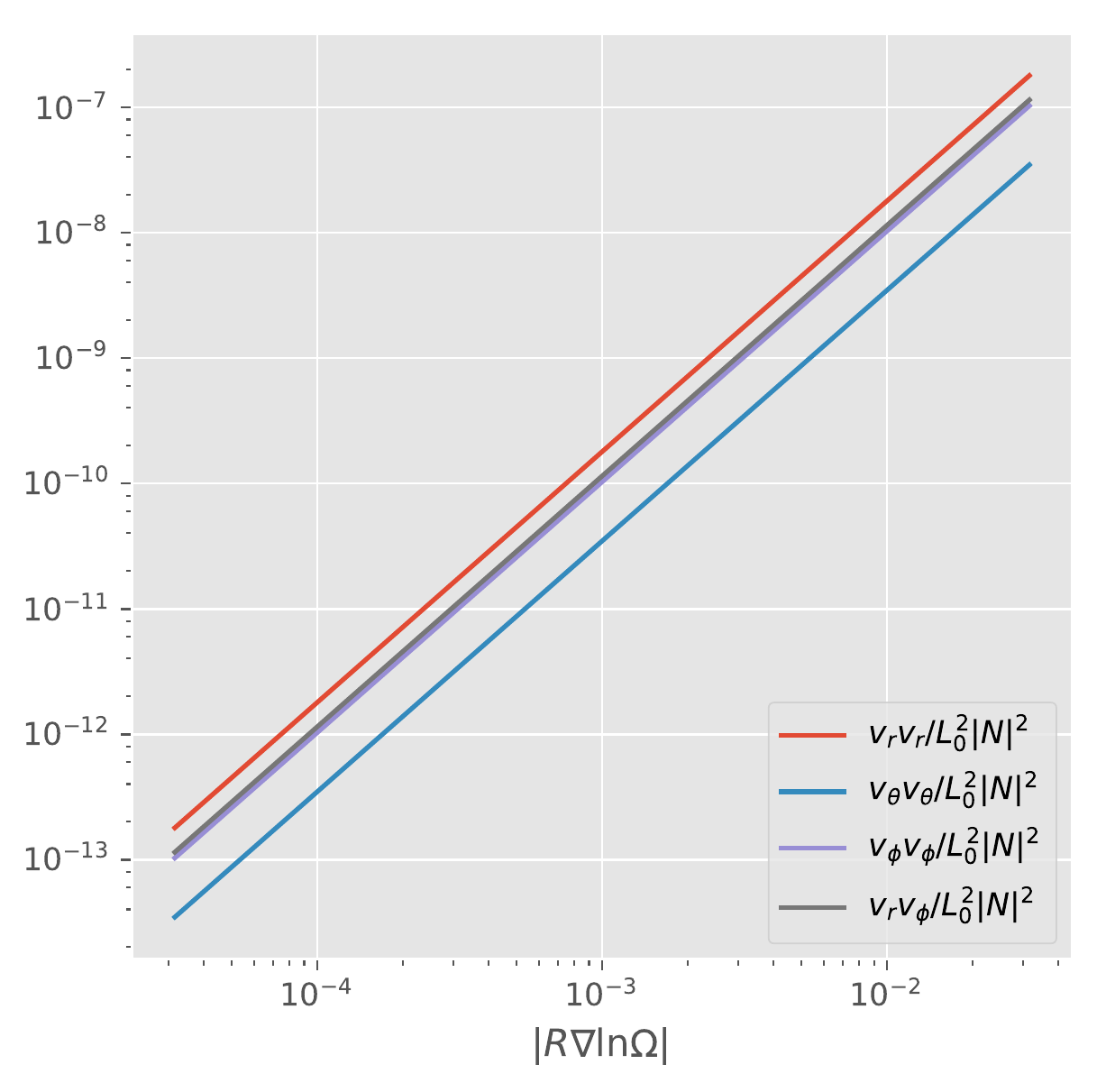}
	\caption{The absolute value of the $r$-$r$ (red), $\theta$-$\theta$ (blue), $\phi$-$\phi$ (purple) and $r$-$\phi$ (grey) velocity correlation functions are shown as a function of $|R\nabla\ln\Omega|$, with both axes log-scaled. These results are for a stably stratified region with differential rotation in the radial direction, $\Omega = 0.1 |N|$ and no magnetic field. The data is computed for a point on the equator with radial differential rotation. All quantities are given in units of the mixing length and the \brvs\ frequency.}
	\label{fig:diffrot3}
\end{figure}

\begin{figure}
	\centering
	\includegraphics[width=\figwidth]{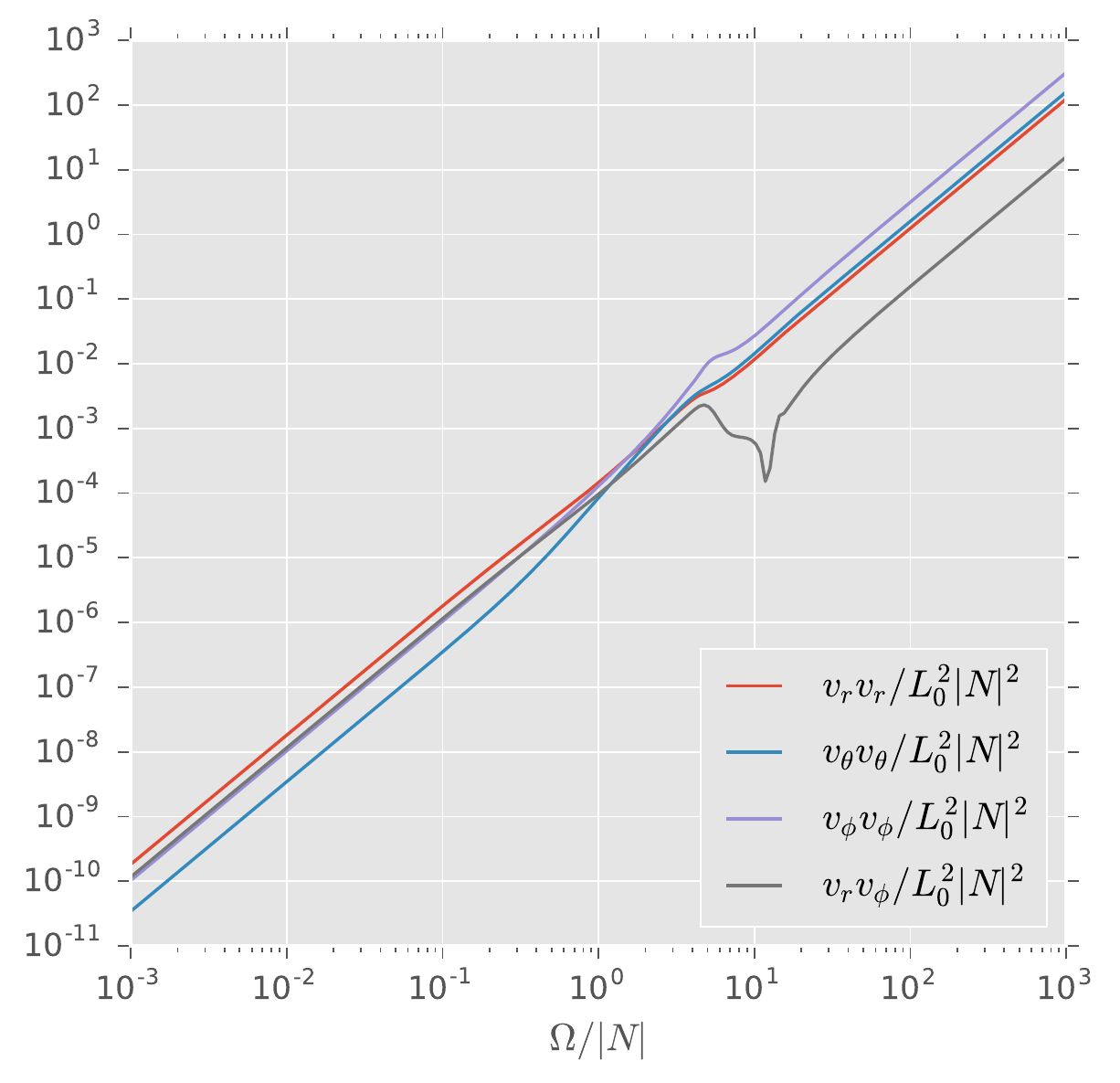}
	\caption{The absolute value of the $r$-$r$ (red), $\theta$-$\theta$ (blue), $\phi$-$\phi$ (purple) and $r$-$\phi$ (grey) velocity correlation functions are shown as a function of $\Omega/|N|$ for fixed $|R\nabla\ln\Omega|=0.1$ with both axes log-scaled. These results are for a stably stratified region with differential rotation in the radial direction and no magnetic field. The data is computed for a point on the equator with radial differential rotation. All quantities are given in units of the mixing length and the \brvs\ frequency. Note that the $r$-$\phi$ and $\theta-\phi$ correlations undergo a sign change at $\Omega/|N| \approx 10 = |R\nabla\ln\Omega|^{-1}$, where terms which are linear in the differential rotation are overtaken by those which are quadratic.}
	\label{fig:diffrot32}
\end{figure}

Fig.\ \ref{fig:diffrot3} shows the dependence on shear strength of all non-vanishing stress components in a rotating stably stratified zone with latitudinal rotational shear.
All exhibit linear scaling with the shear strength.
This is unusual in an otherwise-stable zone because it implies a viscosity which, to leading order, does not depend on the shear.
That is,
\begin{align}
	\nu_{ij} \approx L_0^2 N f_{ij}\left(\frac{\Omega}{|N|}\right),
	\label{eq:nugen}
\end{align}
where $f_ij$ is some function of the angular velocity.
Fig.~\ref{fig:diffrot32} shows the dependence of the stress components on $\Omega/|N|$ for fixed $|R\nabla \ln \Omega|=0.1 $.
The components all scale as $\Omega^2$ in both regimes.
Thus, for instance, $f_{r\phi} = \Omega/|N|$ because the viscosity is the derivative of the stress with respect to the shear, and hence
\begin{align}
	\nu_{r\phi} \approx 10^{-5} L_0^2 \Omega.
	\label{eq:nurt}
\end{align}

The scaling in equation\ \eqref{eq:nurt} arises owing to the centrifugal term, which has a destabilising effect when $\Omega$ increases with $\hat{R}$.
When $|R\nabla\Omega| = 0$ this effect is not present so the system is stable but introducing a small differential rotation produces an acceleration proportional to $\Omega R \boldsymbol{\delta r}\cdot\nabla\Omega$ and hence
\begin{align}
	\partial_t^2 \boldsymbol{\delta r} \approx g^2 \boldsymbol{\delta r} \propto \hat{R}\Omega R \boldsymbol{\delta r}\cdot\nabla\Omega,
	\label{eq:cent}
\end{align}
which means that the stress scales as $\Omega \nabla \Omega$ and thence the viscosity scales as $\Omega$.

\begin{figure}
	\centering
	\includegraphics[width=\figwidth]{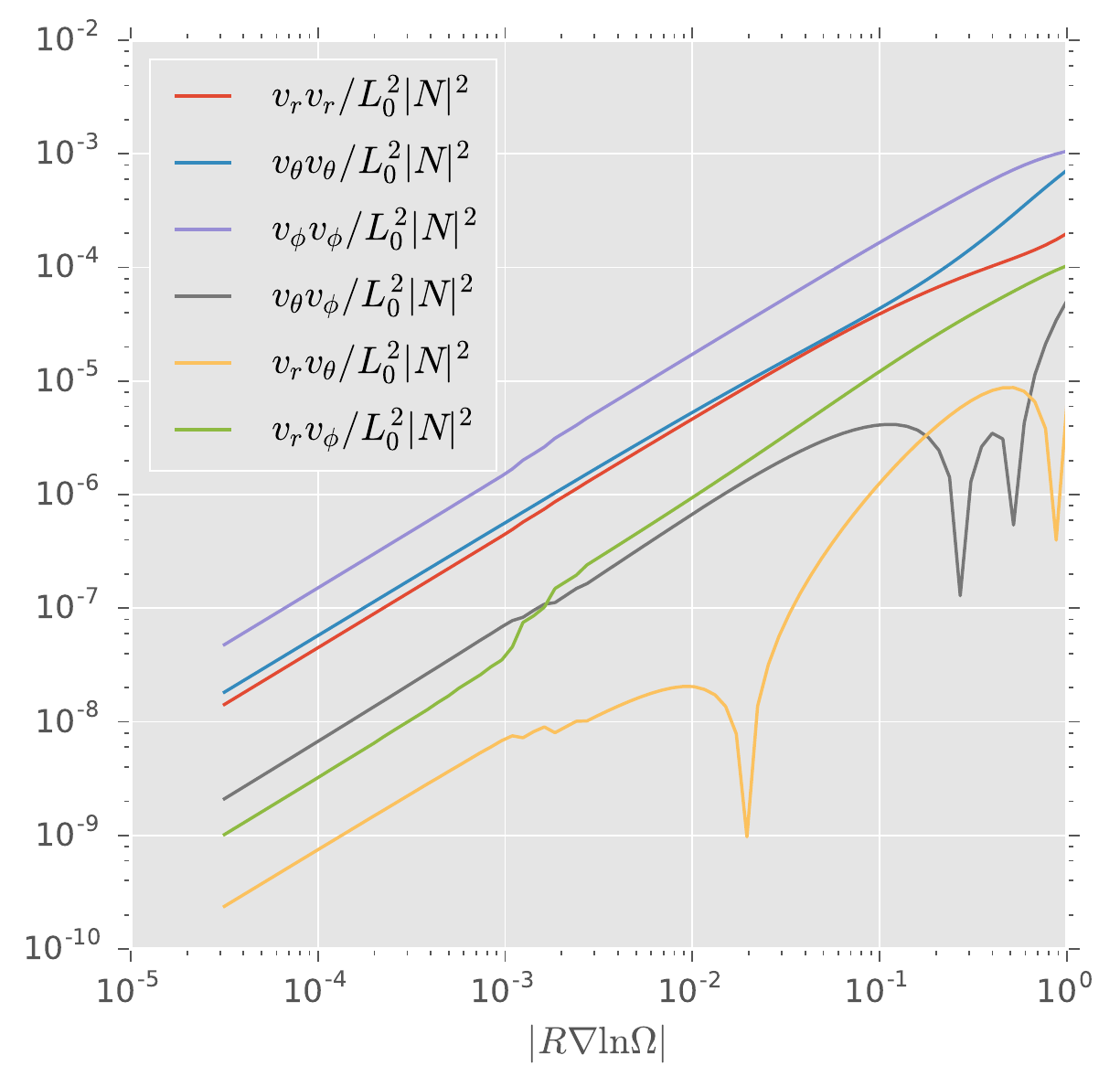}
	\caption{The absolute value of the $r$-$r$ (red), $\theta$-$\theta$ (blue), $\phi$-$\phi$ (purple), $\theta$-$\phi$ (grey), $r$-$\theta$ (yellow) and $r$-$\phi$ (green) velocity correlation functions are shown as a function of $|R\nabla\ln\Omega|$ with both axes log-scaled. The correlation functions are evaluated at first order in the perturbative expansion. These results are for a stably stratified region with differential rotation in the radial direction, $\Omega = 0.1 |N|$ and no magnetic field. The data are computed for a point on the equator with differential rotation at an angle of $\pi/4$. All quantities are given in units of the mixing length and the \brvs\ frequency. Note that the $r$-$\theta$ correlation undergoes a sign change at $|R\nabla\ln\Omega| \approx 10^{-2} \approx (\Omega/|N|)^2$, where terms which are linear in the differential rotation are overtaken by those which are quadratic. Both this and the $\theta-\phi$ component undergo sign changes near $|R\nabla\ln\Omega| = |N|$, where the shear competes directly with the stable stratification.}
	\label{fig:diffrot4}
\end{figure}

\begin{figure}
	\centering
	\includegraphics[width=\figwidth]{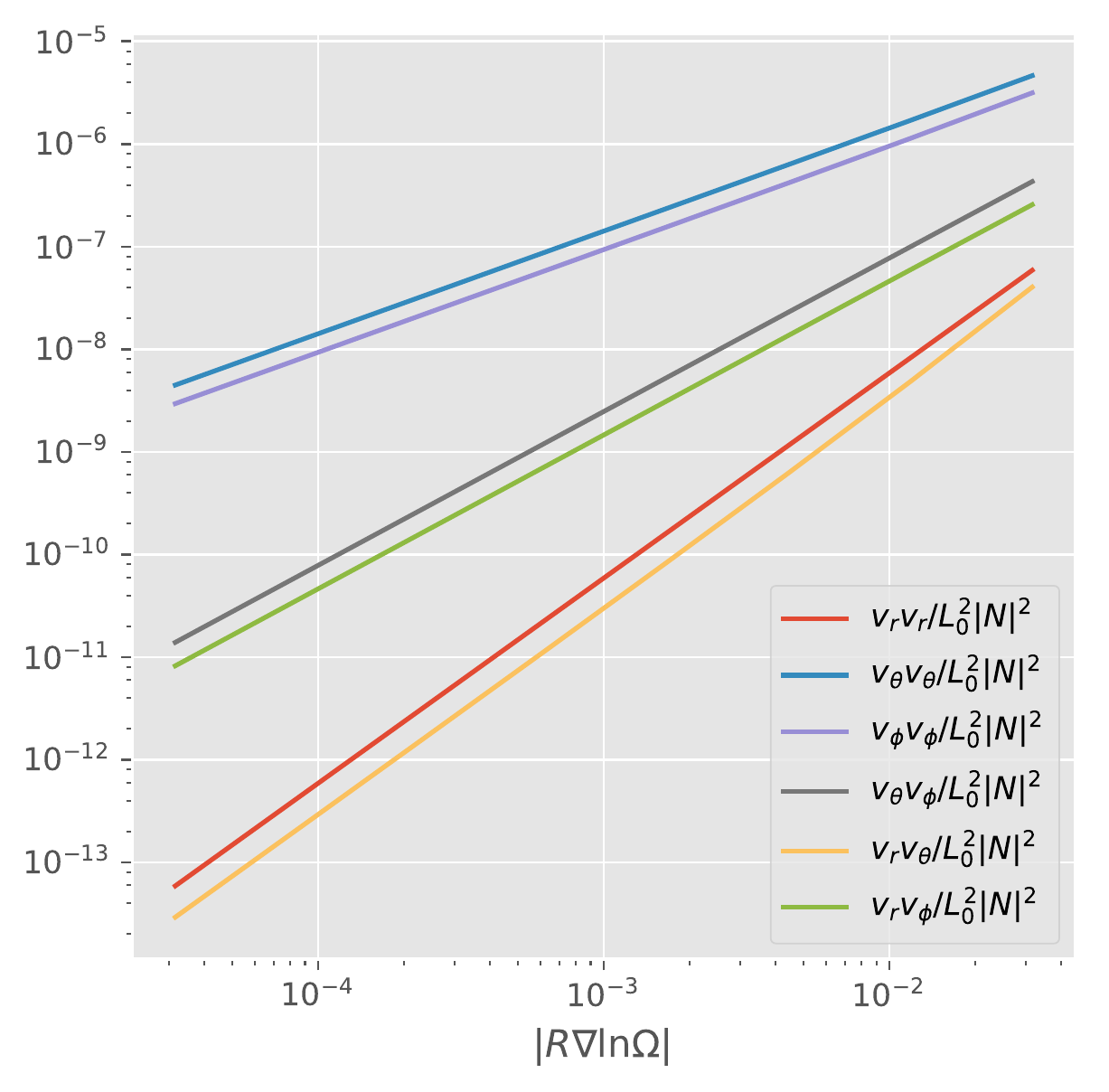}
	\caption{The absolute value of the $r$-$r$ (red), $\theta$-$\theta$ (blue), $\phi$-$\phi$ (purple), $\theta$-$\phi$ (grey), $r$-$\theta$ (yellow) and $r$-$\phi$ (green) velocity correlation functions are shown as a function of $|R\nabla\ln\Omega|$ with both axes log-scaled. The correlation functions are evaluated at zeroth order in the perturbative expansion rather than first order. These results are for a stably stratified region with differential rotation in the radial direction, $\Omega = 0.1 |N|$ and no magnetic field. The data are computed for a point on the equator with differential rotation at an angle of $\pi/4$. All quantities are given in units of the mixing length and the \brvs\ frequency.}
	\label{fig:diffrot40}
\end{figure}

\begin{figure}
	\centering
	\includegraphics[trim = {0 0 0 0}, clip, width=\figwidth]{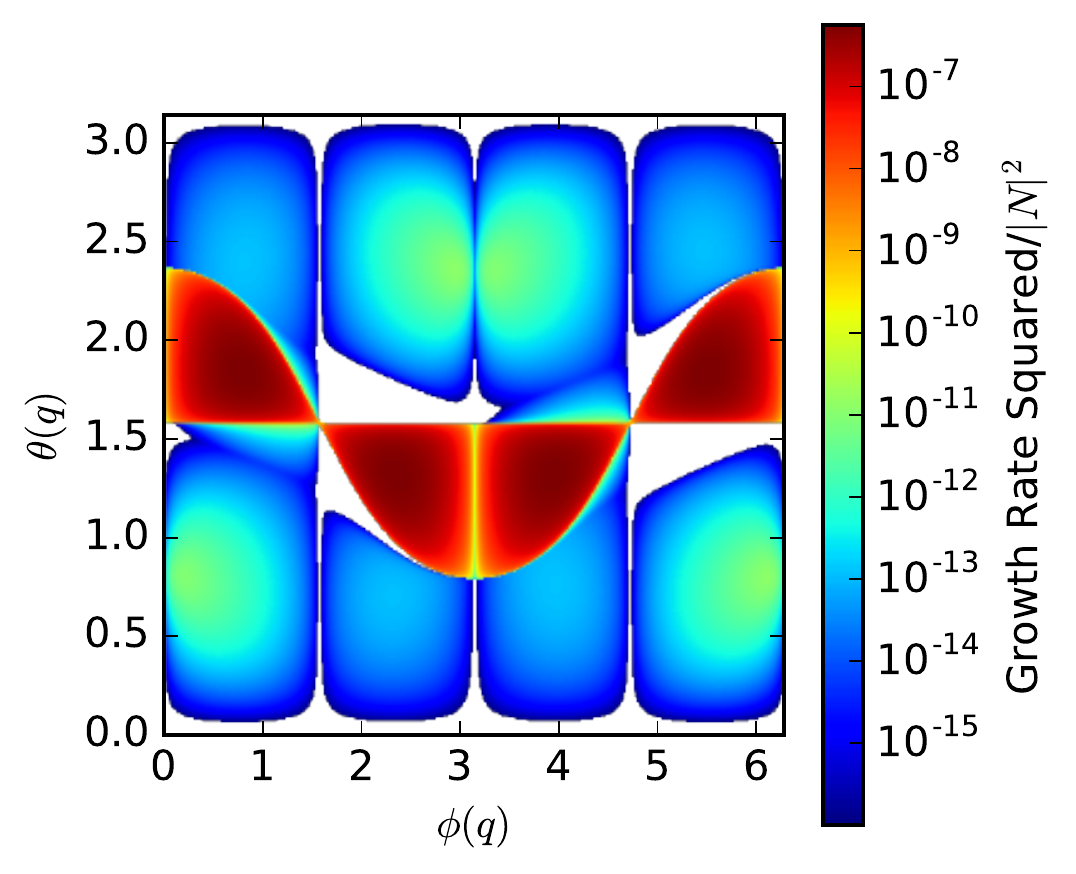}
	\caption{The square of the growth rate is shown as a function of wave-vector orientation on a logarithmic colour scale. The wave-vector is specified by a magnitude and two angles, $\theta(q)$ and $\phi(q)$, which are spherical angles relative to the $\hat{z}$ direction. These rates were computed with a zeroth-order expansion. Regions with squared growth rates below $10^{-16}$ are shown in white. All quantities are given in units of the mixing length and the \brvs\ frequency.}
	\label{fig:lobe0}
\end{figure}

To better understand the effect of our perturbative corrections we computed the same results without them.
This produced stresses which were zero to within numerical precision in all cases, indicating that the entire contribution in this case is coming from the perturbation.
However with a different angle of differential rotation we obtained non-zero results.
It is instructive then to compare Fig.~\ref{fig:diffrot4} with Fig.~\ref{fig:diffrot40}.
These show the same correlation functions as each other in the same physical scenario, with differential rotation this time at an angle of $\pi/4$, but the former uses the first order perturbative expansion while the latter only expands to zeroth order.
The difference between the two calculations is striking: many of the correlation functions have fundamentally different scalings when the perturbative corrections are taken into account.
In particular the non-vanishing stresses are quadratic in the shear, whereas they are all linear in the shear in the expanded calculation.
This difference relates in part to the centrifugal term, which couples the displacement to the acceleration.
Without expanding the equations of motion we would have $\delta r \propto \delta v$, because the mode would need to be an eigenvector of $\mathsf{M}$.
The modes which couple to the centrifugal term would still grow according to equation\ \eqref{eq:cent} but, for most modes, arranging for the displacement to couple to this term requires coupling to the stabilising buoyant term too.
To make this clearer, in Fig.\ \ref{fig:lobe0} we have computed the growth rate as a function of wave-vector orientation without using the perturbative expansion.
There are several rapidly-growing regions, oriented at angles of $\pm \pi/4$ relative to the vertical.
These angles represent a compromise between maximising the magnitude of the centrifugal acceleration and maximising its projection on to the velocity, both subject to the Boussinesq condition that motion be in the plane perpendicular to $\boldsymbol{q}$.

\begin{figure}
	\centering
	\includegraphics[trim = {0 0 0 0}, clip, width=\figwidth]{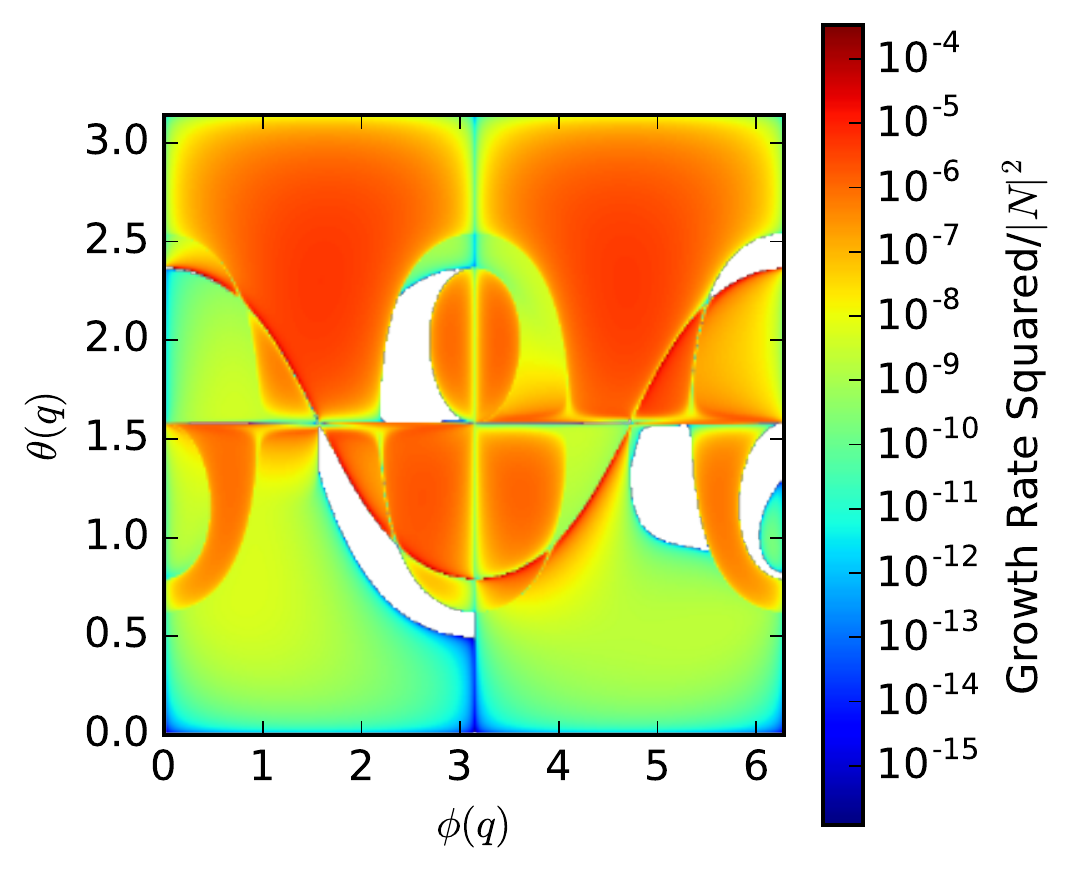}
	\caption{The square of the growth rate is shown as a function of wave-vector orientation on a logarithmic colour scale.  The wave-vector is specified by a magnitude and two angles, $\theta(q)$ and $\phi(q)$, which are spherical angles relative to the $\hat{z}$ direction. These rates were computed with a first-order expansion. Regions with squared growth rates below $10^{-16}$ are shown in white. All quantities are given in units of the mixing length and the \brvs\ frequency.}
	\label{fig:lobe1}
\end{figure}

By contrast the growth rates in the expanded system, shown in Fig.\ \ref{fig:lobe1}, are significant over a much wider swath of parameter space.
This is because, in the expanded system, the displacement and velocity need not be parallel so the displacement can be chosen to maximise the centrifugal term while the velocity can be chosen to maximise the projection of the acceleration on to the velocity.

\subsection{Baroclinic Instability}

\begin{figure}
	\centering
	\includegraphics[width=\figwidth]{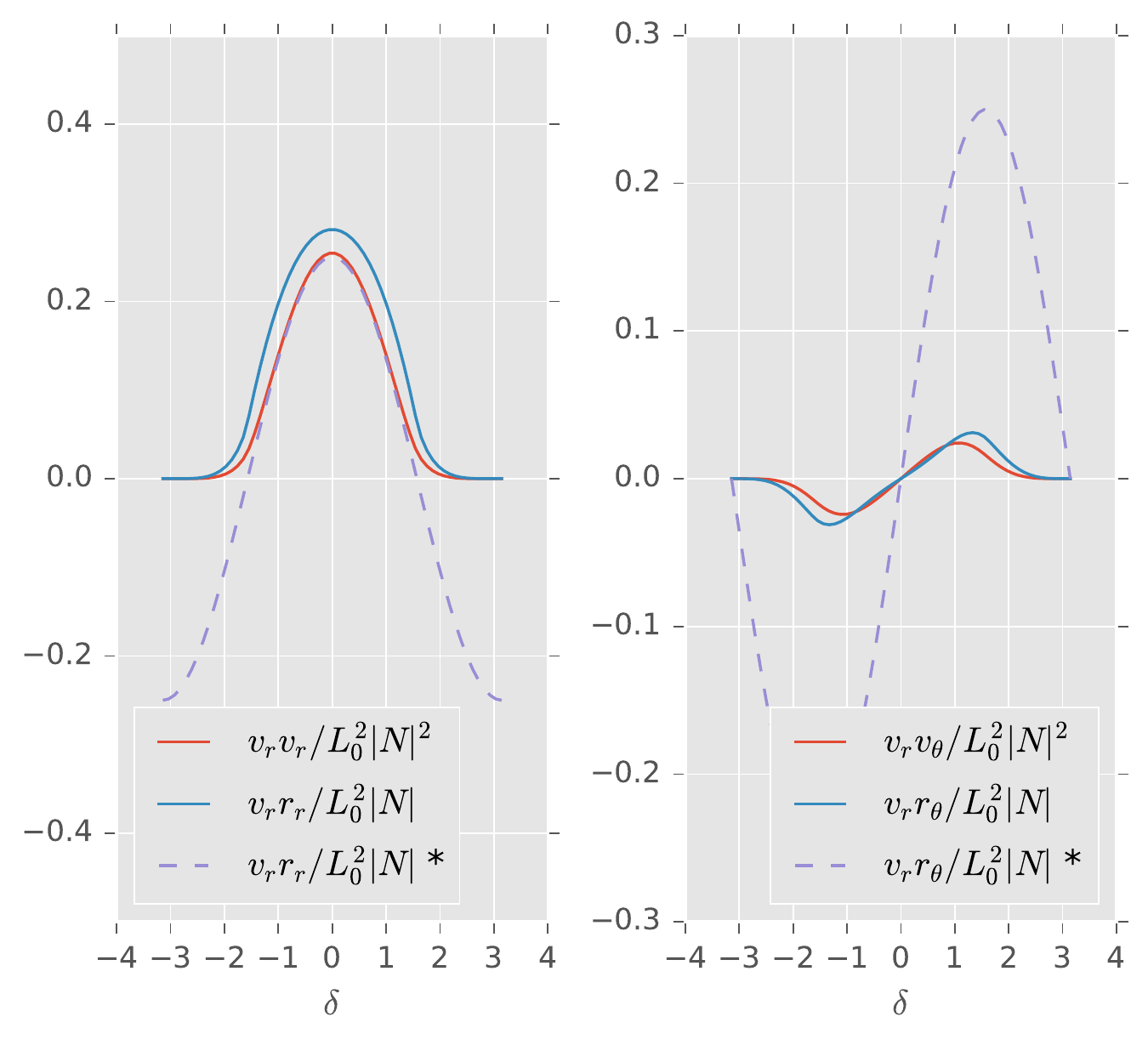}
	\caption{Various correlation functions are shown as a function of the angle $\delta$ between the entropy and pressure gradients. The functions are the $r-r$ (left) and $r-\theta$ (right) velocity (red) and diffusivity (blue) correlation functions. These results are for a non-rotating convective region on the equator and with no magnetic field. Shown in purple (*, dashed) for comparison is the KR result. This agrees in sign, and for $r-r$ agrees in scale, but their $r-\theta$ prediction is considerably larger. Notably this comparison is precisely as $\cos(\delta)$ (left) and $\sin(\delta)$ (right) and crosses zero at non-extremal angles. This is most likely because their theory is not designed for nearly-stable regions with extreme baroclinicity. All quantities are given in units of the mixing length and the \brvs\ frequency.}
	\label{fig:baro}
\end{figure}

The baroclinic instability arises in otherwise stably stratified fluids when the entropy gradient is not parallel to the pressure gradient\ \citep{1980DyAtO...4..143K}.
In fact this is part of a family of instabilities which includes the convective instability\ \citep{1965ApJ...142.1257L}.
This family provides a continuous connection between the unstable convective and stably stratified limits.
To explore it consider Fig.\ \ref{fig:baro} which shows the variation of $r-r$ and $r-\theta$ correlation functions against the angle $\delta$ between the entropy gradient and the pressure gradient.
The radial correlations peak when the two gradients are aligned.
This is the convective limit.
These correlations fall to zero in the opposing limit where the two gradients are anti-aligned, which is the stably stratified limit.
In between these limits the behaviour is approximately that of $\cos^2 \delta$.

By contrast the $r-\theta$ correlations behave approximately as $\sin \delta$, and vanishes when $\delta = 0$.
This is because both the aligned and the anti-aligned limits are spherically symmetric and so must have this correlation function vanish.
Deviations from the convective limit give rise to linear scaling so the convective baroclinic instability transports heat and momentum at first order in the baroclinicity.
This is an entirely distinct phenomenon from the thermal wind balance, which is a large-scale effect while this results from integrating out the small-scale turbulent modes.
In the stable limit perturbations arise quadratically, a deviation from the behaviour of $\sin \delta$.
This is because there are no existing turbulent motions to perturb, and so each position and velocity component is linear in $\delta$ and gives rise to a quadratic two-point correlation function.

\subsection{Stellar Magnetism}

\begin{figure}
	\centering
	\includegraphics[width=\figwidth]{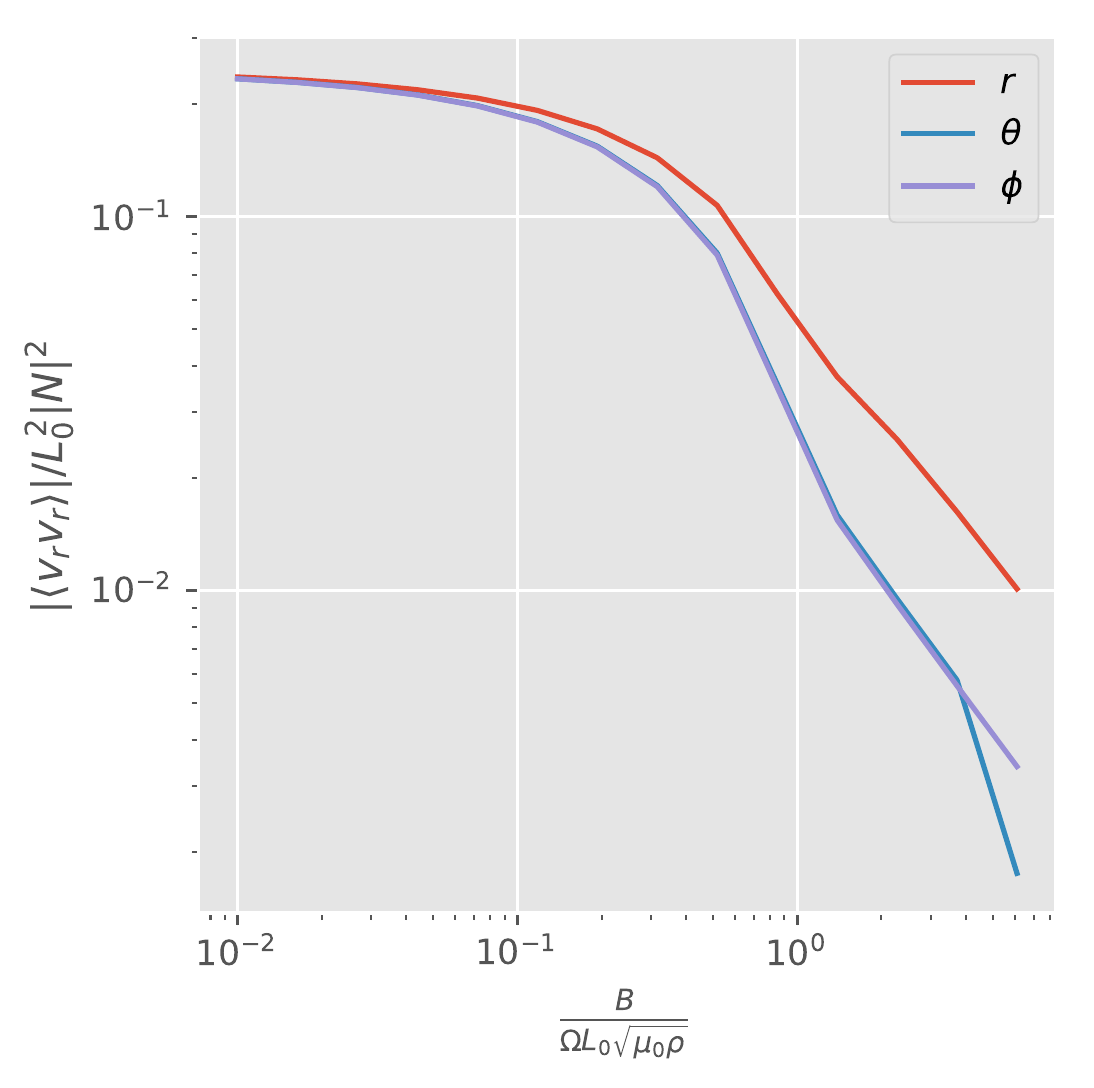}
	\caption{The stress $\langle \delta v_r \delta v_r \rangle$ is shown as a function of magnetic field strength. The magnetic field is polarised radially (red), longitudinally (purple) and latitudinally (blue). The system is rigidly rotating at $\Omega=0.1|N|$ at a latitude of $\pi/4$. All quantities are given in units of the mixing length and \brvs\ frequency.}
	\label{fig:polRR}
\end{figure}

We now turn to the impact of the magnetic field on convective turbulence in stars.
Fig.\ \ref{fig:polRR} shows $\langle \delta v_r \delta v_r \rangle$ in a mildly rotating ($\Omega=0.1|N|$) convection zone as a function of $B$ for three polarisations; radial ($\boldsymbol{B} \parallel \hat{\boldsymbol{r}}$), latitudinal ($\boldsymbol{B} \parallel \hat{\boldsymbol{\theta}}$),  and longitudinal ($\boldsymbol{B} \parallel \hat{\boldsymbol{\phi}}$).
As the field increases the stress falls off.
This is because the field quenches the turbulence by providing a stabilising restoring force, and is in general agreement with~\citet{1986A&A...168...89C}.
Interestingly the only significant differences are between the radial and angular field polarisations!
The $\theta$ and $\phi$ polarisations show precisely the same behaviour out to very strong fields.
This is a result of symmetry, because the radial stress is not sensitive to rotation about the radial direction.
The deviation seen with strong fields is a numeric artefact and decreases with increasing integration time.

\begin{figure}
	\centering
	\includegraphics[width=\figwidth]{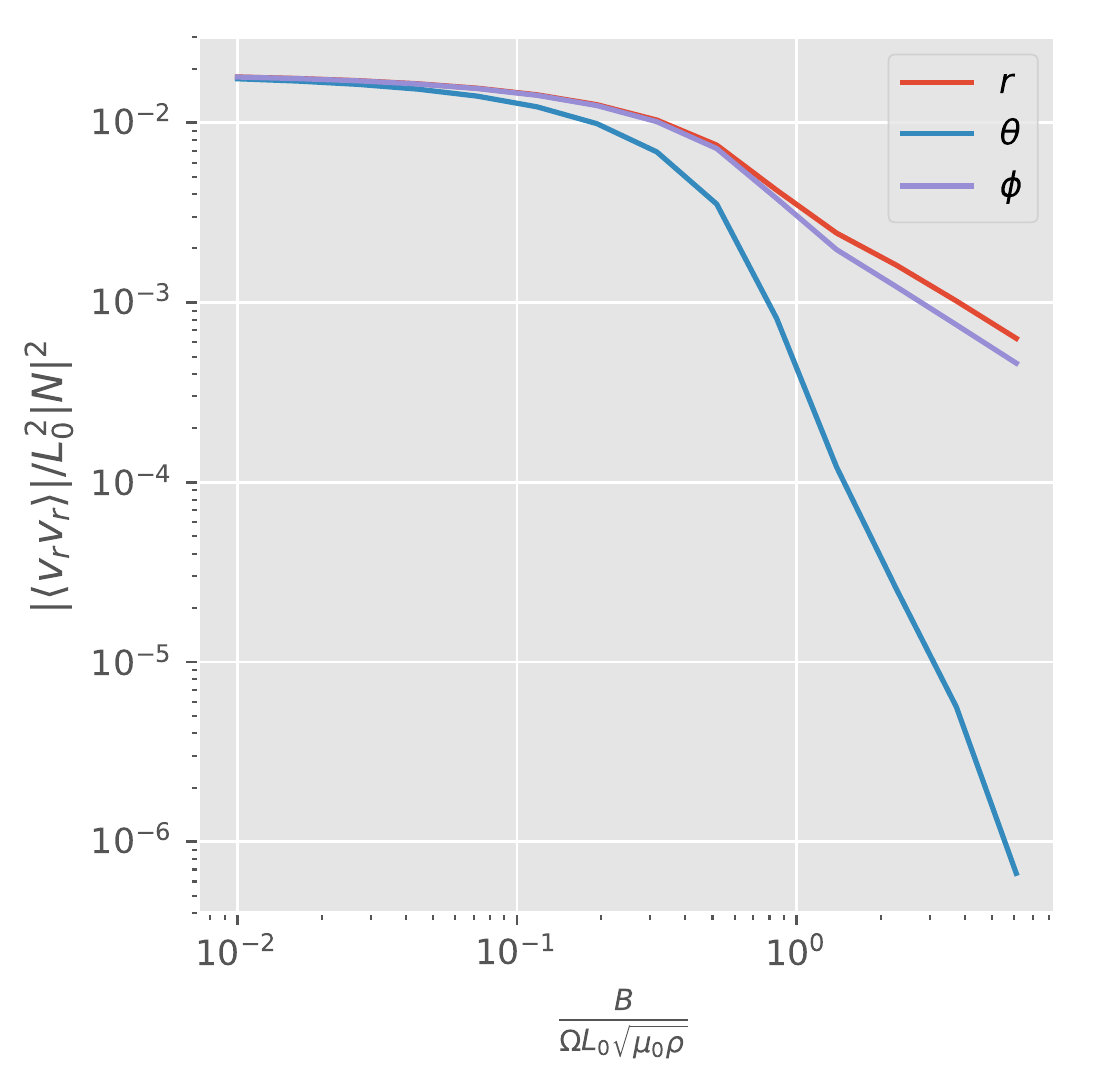}
	\caption{The stress $\langle \delta v_r \delta v_\phi \rangle$ is shown as a function of magnetic field strength. The magnetic field is polarised radially (red), longitudinally (purple) and latitudinally (blue). The system is rigidly rotating at $\Omega=0.1|N|$ at a latitude of $\pi/4$. All quantities are given in units of the mixing length and \brvs\ frequency.}
	\label{fig:polRP}
\end{figure}

By contrast consider $\langle \delta v_r \delta v_\phi \rangle$, shown in Fig.\ \ref{fig:polRP}.
This component, along with the corresponding Maxwell stress, is responsible for transporting angular momentum.
Interestingly it shows differences amongst all polarisations, with the strongest difference between the $\theta$ polarisation and the others.
This is because the stress is mixed between different directions and so is sensitive to all variations in the magnetic field direction.
The large difference of the $\theta$ polarisation relative to the others reflects the fact that motion is damped perpendicular to the magnetic field so the $\theta$ polarisation damps motion in both directions involved in this component of the stress whereas the $r$ and $\phi$ polarisations only damp motion in one of the two directions.

\subsection{Magnetorotational Instability}

As a final example we consider the magnetorotational instability\ \citep{1960PNAS...46..253C}.
This instability arises in magnetised fluids undergoing Keplerian orbital motion.

\begin{figure}
	\centering
	\includegraphics[width=\figwidth]{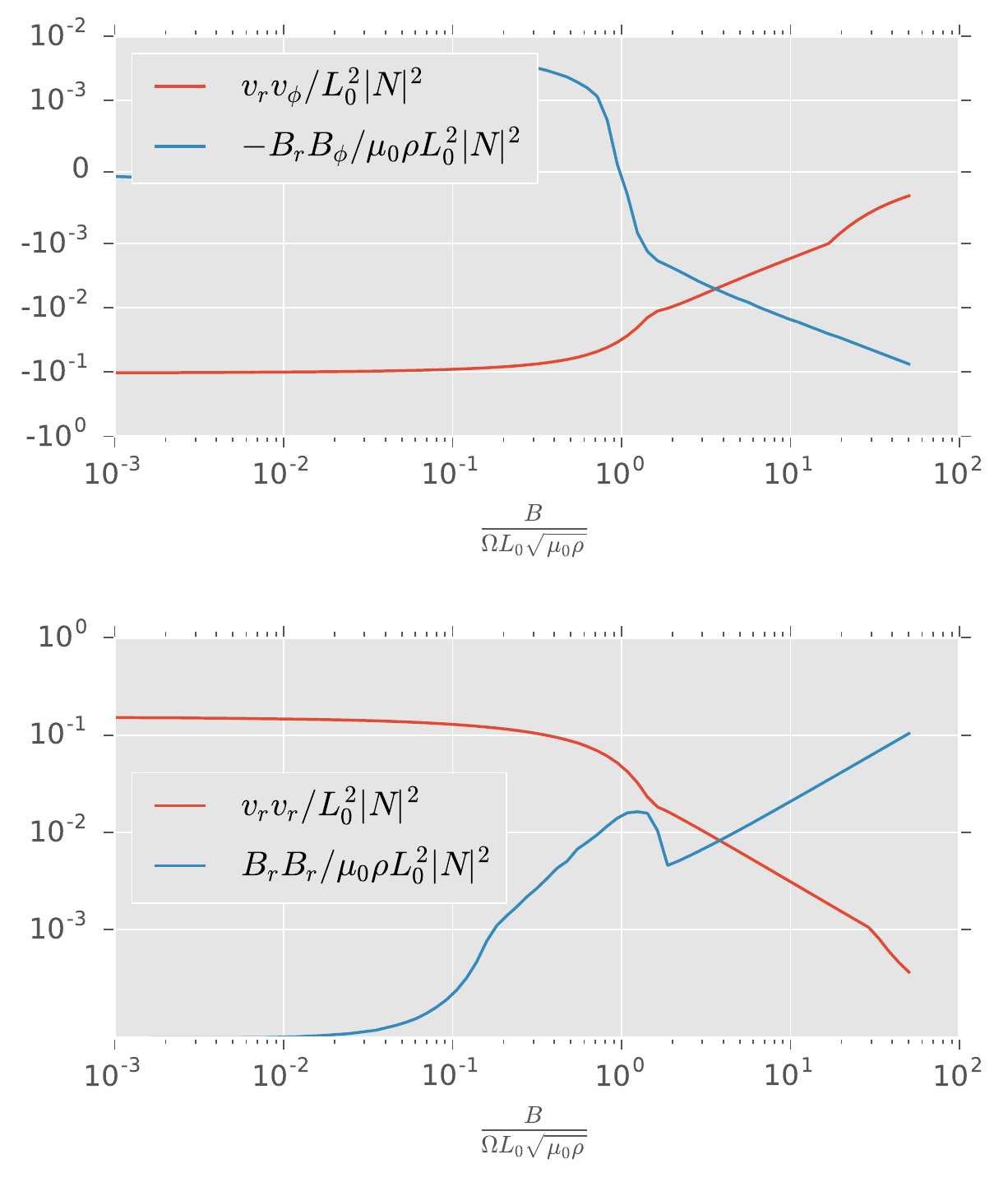}
	\caption{
	The $r-r$ (top) and $r-\phi$ (bottom) are shown as a function of $B$ for a Keplerian disc.
	Reynolds (velocity) stresses are in red and the Maxwell (magnetic) stresses are in blue.
	Note that it is the negative $r-r$ Maxwell stress which is shown to make the comparison with the Reynolds stresses clearer. 
	The magnetic field is taken parallel to $\hat{z}$. The system is taken to be stably stratified in the vertical direction with $|N| = \Omega$ and hence $L_0 = h = R$. All quantities are given in units of the mixing length and $\Omega$.}
	\label{fig:mri}
\end{figure}

Fig.\ \ref{fig:mri} shows the $r-r$ and $r-\phi$ Reynolds and Maxwell stresses for an accretion disc with a vertical magnetic field.
Contrary to predictions\ \citep{1960PNAS...46..253C} not all of the Reynolds stresses vanish in the zero-field limit.
This is because the linear system supports short-term growing modes but, while they only grow in the short-time limit, our numerical methods are not sensitive to that effect at this order.
In principle, at higher order, this phenomenon should become evident and so this may be interpreted as an artefact associated with our expanding to low order in $|R\nabla\Omega| > 1$.
Despite this, it is likely that other non-magnetic processes can destabilise these modes even in the long term and so we feel it is appropriate to at least consider them~\citep[cf.][]{math/0607282}.
The Maxwell stresses by contrast do vanish as $B\rightarrow 0$.
This is to be expected because they are proportional to $B^2$.
Additionally, the $\theta$-$\phi$ stresses vanish for all $B$ because the system is symmetric under reversing both $\theta$ and $\phi$, and similarly the $r$-$\theta$ stresses vanish because the system is symmetric under the simultaneous reversal of $r$ and $\theta$.

As the magnetic field increases the $r-\phi$ Reynolds stress changes sign.
This indicates the onset of MRI modes, which have the opposite sign to the zero-field correlations.
This effect saturates when $v_A \approx \Omega h$, where $h$ is the scale height of the disc.
The total $r-\phi$ stress saturates at roughly $10^{-2} (h \Omega)^2$, which lies between those typically found in simulations and those inferred from observations\ \citep{Starling2004, 2007MNRAS.376.1740K}.
Note that at the saturation point the Maxwell and Reynolds stresses are comparable, and beyond this point the Maxwell stress increases while the Reynolds stress falls off.

Above the saturation point the Reynolds stresses drop off as the magnetic field quenches the turbulence.
This is precisely what is expected for the MRI\ \citep{1991ApJ...376..214B}.
The Maxwell stresses, however, continue to grow, again in line with expectations.
Some care is required to interpret these results because they were computed for a fixed field and that field may or may not be stable under the action of the turbulence it generates~\citep{2006MNRAS.372..183P}.
Furthermore there are challenges with the $\alpha$-disk prescription which make the specific stress components more difficult to interpret~\citep{2008MNRAS.383..683P}.
Nevertheless it is encouraging that what we see matches well with both observations and simulations.

\section{Conclusion}
\label{sec:conclusion}

We have derived a turbulent closure model which incorporates shear, rotation and magnetism as well as a full three-dimensional spectrum of fluctuations.
We have also presented a new perturbative approach to incorporate time-dependence in the evolution equations.
This model, which is implemented in an open source numerical software package, fully reproduces many known phenomena such as the MRI, baroclinic instability, rotational quenching and more classic shear instabilities.

Using this model we have determined the asymptotic behaviour of a wide variety of correlation functions and transport coefficients under a wide range of circumstances, many of which do not appear in the literature.
We have further explored the behaviour of turbulent transport coefficients in intermediate regimes where no single phenomenon dominates, such as in the critical MRI.
In these cases the behaviour is generally complex and does not separate easily into components associated with the different pieces of input physics.

The closure formalism developed here fills a new niche in the landscape of solutions to turbulent transport, covering enough phenomena to be useful to understand those operating in stars, planets and accretion discs, while being rapid enough to be incorporated into stellar evolution codes on nuclear timescales.

In the future we hope to provide further refinements and comparisons with direct numerical simulations as well as experiments.
In addition, it would be interesting to explore the results of this model to higher order in the shear and, even at this order, there are many results which deserve more analysis than we have given here.

\section*{Acknowledgements}

ASJ acknowledges financial support from a Marshall Scholarship as well as support from the IOA, ENS and CEBS to work at ENS Paris and CEBS in Mumbai.
PL acknowledges travel support from the french PNPS (Programme National de Physique Stellaire) and from CEBS.
CAT thanks Churchill College for his fellowship.
SMC is grateful to the IOA for support and hopsitality and thanks the Cambridge-Hamied exchange program for financial support.
The authors also thank Rob Izzard and STFC Grant ST/L003910/1 for CPU cycles which aided in this work.




\bibliographystyle{mnras}
\bibliography{mlt.bib} 



\appendix

\section{Turbulent Index}
\label{appen:n}

The general question of  which turbulent index to use and under what circumstances remains open though many specific cases are well understood.
In the case of isotropic incompressible turbulence the Kolmogorov index is well-known to be $n=11/6$\ \citep{1941DoSSR..30..301K}.
There is more debate over the index to use for convection, with answers ranging from $n=5/2$\ \citep{1994EL.....25..341B} to $n=21/10$\ \citep{1989PhRvL..62.2128P} and $n=2.4\pm 0.2$\ \citep{1999PhRvL..83.4760A}.
There has also been work attempting to determine the spectrum in a context-sensitive manner through energy balance arguments~\citep{1986JSCom...1....3Y}.
In the magnetised case sources differ even more, with some suggesting that this range still applies\ \citep{PhysRevLett.45.144}, some arguing for a Kolmogorov-like spectrum\ \citep{1995ApJ...438..763G} and others giving a range of indices depending on geometry and the direction of the wavevector\ \citep{1994ApJ...432..612S}.

From numerical experiments with our closure model we have found that the magnetic stress scales sufficiently rapidly with $k$ that it is divergent for $n=11/6$ and not for $n=8/3$.
This favours the scenario of \citet{1995ApJ...438..763G}, who argue that in the strongly-magnetised limit the index ought to be $n=8/3$.

In order to consistently treat both the non-magnetic and the strongly-magnetised limits, we choose a simple prescription in which $n=11/6$ when one of $|N|$, or $|R\nabla\Omega|$ exceeds $k v_A$ and use $n=8/3$ otherwise.
This means that there is a critical wavenumber
\begin{align}
	k_c \equiv \frac{\max\left(|N|, |R\nabla\Omega|\right)}{v_A}
\end{align}
at which the spectrum changes.
In the non-magnetic case the evolution matrix is independent of the magnitude of the wavevector and so altering the index just alters the correlation coefficients by a multiplicative factor.
In the magnetic case the potential for error is larger because the magnitude of the wavevector is relevant but there appears to be no consensus on the best prescription and so we make do with what is available.

\section{Boussinesq Oddities}
\label{appen:b}

In this work we have taken the Boussinesq approximation.
In Fourier space this is
\begin{equation}
\boldsymbol{q}\cdot\tilde{\delta\boldsymbol{r}}=0.
\end{equation}
Taking the time derivative of both sides we see that
\begin{equation}
\partial_t\left(\boldsymbol{q}\cdot \delta\tilde{\boldsymbol{r}}\right) = \boldsymbol{q}\cdot\delta\tilde{\boldsymbol{v}} + \delta\tilde{\boldsymbol{r}}\cdot\partial_t\boldsymbol{q} = 0.
\end{equation}
As a result
\begin{equation}
	\delta \tilde{\boldsymbol{v}}\cdot\boldsymbol{q} = - \delta\tilde{\boldsymbol{r}}\cdot\partial_t\boldsymbol{q} \neq 0.
\end{equation}
This is quite peculiar, but is just an artefact of our coordinate system.
Because the wavevectors are time-dependent, maintaining the volume of a fluid parcel requires that the displacement be orthogonal to the wavevector, which actually means that the velocity is generally not orthogonal to the wavevector.

\section{Software Details}
\label{appen:software}

The software used for this work is Mixer version 1, which we have released under a GPLv3 license at \url{github.com/adamjermyn/Mixer}.
All data produced for this work are available at the same location as HDF5 tables with attributes documenting the physical inputs.
Post-processing and visualisation of the data was with the Python modules Numpy\ \citep{5725236} and Matplotlib\ \citep{4160265} and the relevant scripts for this are included with Mixer.

The core of Mixer is written in C++, for performance reasons, and the code is supplied with a Makefile which supports compilation on both Linux and MacOS.
Mixer makes use of the Eigen library \citep{eigenweb} for linear algebra.
Mixer also uses the Cubature library for numerical integration.
This library is an implementation of the algorithms by \citet{GENZ1980295} and \citet{Berntsen:1991:AAA:210232.210233}.
These integration routines are supplemented by a Python integration routine tailored for integrands with small support regions.
The details will be explored in later work.
In addition, many routines provide a Python interface.
Currently Mixer only supports single-threaded operation, though it may be used inside parallelised scripts through the Python wrapper.
The version of Mixer used to generate the data in this work was compiled against Cubature version 1.0.2 and Eigen version 3.3.3, though the code does not use any features which require recent versions, so many likely suffice.

Mixer is optimised for convecting systems for which achieving accuracy better than $10^{-5}$ relative and absolute typically requires between $1\mathrm{ms}$ and $1\mathrm{s}$ on a single core of a 2016 Intel CPU.
This is further improved when the differential rotation is minimal, in which case the perturbative expansion may be turned off to save a factor of several in runtime.
In stably stratified zones and those with magnetic fields up to $10^3\mathrm{s}$ may be required to achieve good convergence.

In cases where the code has more difficulty it is quite likely that Mixer becomes the bottleneck in simulations and so, under these circumstances, we recommend tabulating results in advance.
This is still considerably more performant than direct numerical simulation, and the results can generally be guaranteed to converge at much higher precision, so that derivatives may be extracted as well.

At various points in the software we must divide by the magnitude of the velocity of an eigenmode.
This may approach zero in some cases.
To avoid dividing by zero in these cases we place a lower bound on this magnitude, such that
\begin{align}
	|\delta v|^2 \geq \epsilon,
\end{align}
where $\epsilon = 10^{-20} L_0^2 |N|^2$ in the calculations presented in this work.
This corresponds to setting an upper bound on the length scale $d$ of the displacements $\delta\boldsymbol{r}$, namely
\begin{align}
	|\delta r|^2 \leq L_0^3 |N| \epsilon^{-1/2},
\end{align}
which means that $d=10^{10} L_0$ in this work.

\begin{figure}
	\centering
	\includegraphics[width=\figwidth]{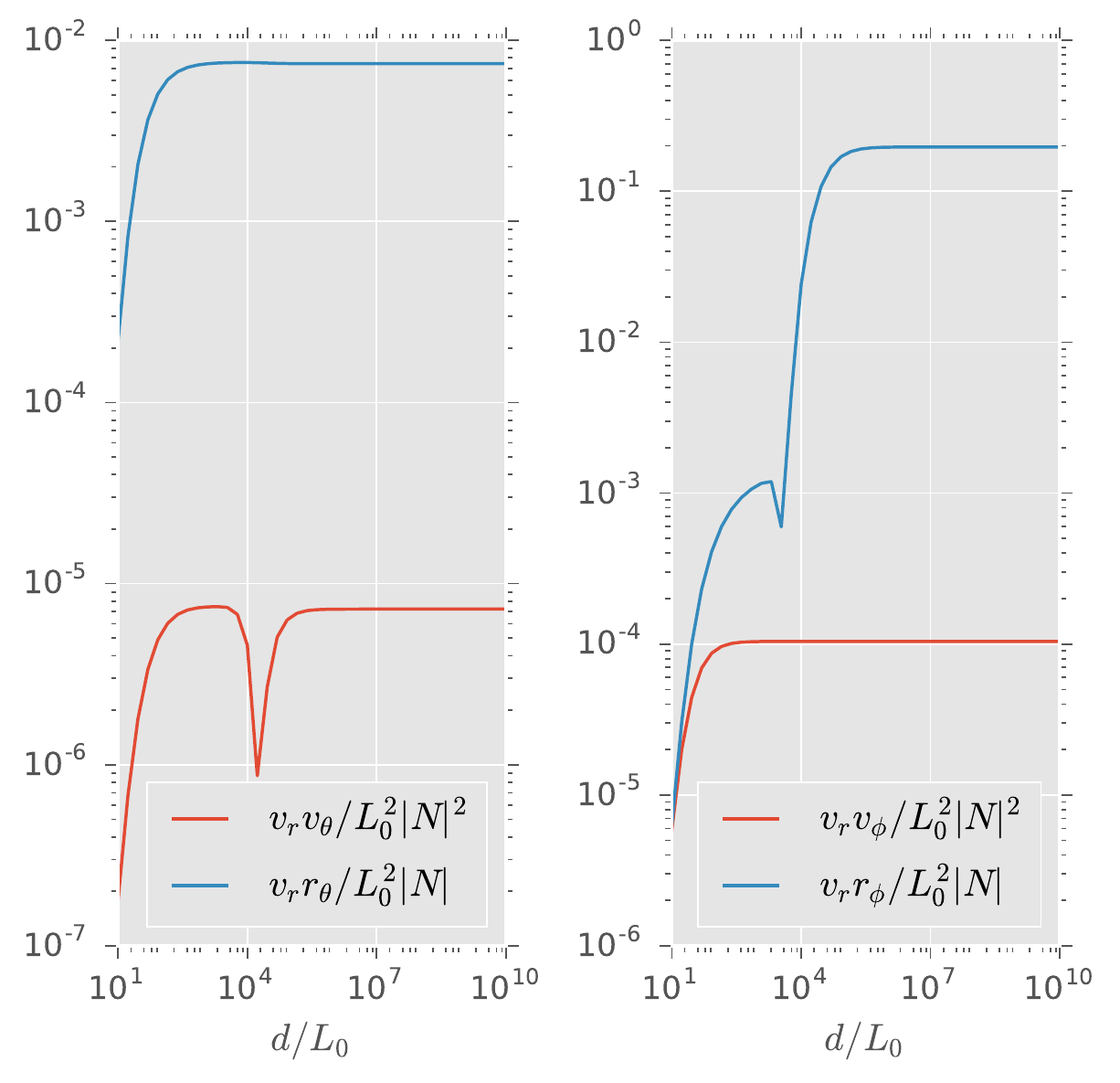}
	\caption{The absolute values of $\langle \delta v_r \delta v_\theta \rangle$ (left) and $\langle \delta v_r \delta v_\phi \rangle$ (right) are shown as functions of $d$, with both axes log-scaled. These results are for a stably stratified region with differential rotation in the radial direction with $|R\nabla\ln\Omega| = 10^{-3}$, $\Omega = 0.1 |N|$ and no magnetic field. The data is computed for a point on the equator with differential rotation at an angle of $\pi/4$. All quantities are given in units of the mixing length and the \brvs\ frequency.}
	\label{fig:eps}
\end{figure}

To verify that this numerical fix does not impact our results we have examined the correlation functions in several scenarios as a function of this numerical cutoff $L$.
For example, figure\ \ref{fig:eps} shows the $r-\theta$ and $r-\phi$ correlations as functions of $d$ for a stably stratified differentially rotating system.
The results are constant over many orders of magnitude so long as $d > 10^5 L_0$, which is easily satisfied by our default.

\bsp	
\label{lastpage}
\end{document}